\documentclass[%
 reprint,
superscriptaddress,
 amsmath,amssymb,
 aps,
]{revtex4-1}

\usepackage{graphicx}
\usepackage{dcolumn}
\usepackage{bm}

\usepackage{braket}
\usepackage{url}
\usepackage[capitalize]{cleveref}
\usepackage{xcolor}
\usepackage{amssymb}

\newcommand{\alice}{\mathbf{A}}
\newcommand{\bob}{\mathbf{B}}
\newcommand{\eve}{\mathbf{E}}

\newcommand{\cowsig}[1]{\ket{\varphi_{#1}}}
\newcommand{\vacsig}{\cowsig{\textrm{vac}}}

\newcommand{\detector}[1]{ {D_{#1}} }

\newcommand{\detbdatalabel}{ {\textrm{D}} }
\newcommand{\detbmonitlabel}[1][]{ {\textrm{M}{#1}} }
\newcommand{\detevelabel}[1][]{ {\textrm{E}{#1}} }

\newcommand{\detbdata}{ \detector{\detbdatalabel} }
\newcommand{\detbmonit}[1][]{ \detector{\detbmonitlabel[#1]} }
\newcommand{\deteve}[1][]{ \detector{\detevelabel[#1]} }

\newcommand{\detectordata}{{\detbdatalabel}}
\newcommand{\detectormonit}[1]{{\detbmonitlabel[#1]}}
\newcommand{\detectoreve}[1][]{\detevelabel[#1]}

\newcommand{\effeve}{{ \eta_E }}
\newcommand{\effbob}{{ \eta_B }}
\newcommand{\darkcount}[1]{ p_{\textrm{d}}^{#1} }

\newcommand{\transmittance}{ \eta_{\rm ch} }
\newcommand{\attconstant}{ \alpha_{\rm ch} }

\newcommand{\attackratio}{ \tau_a }
\newcommand{\attackedlabel}{ \text{a} }
\newcommand{\legitlabel}{ {\bar{\text{a}}} }

\newcommand{\eveinput}[1][]{ \chi_{#1} }

\newcommand{\amperror}{\delta}

\newcommand{\intenseve}[1][]{ \mu_{\detevelabel[#1]|\eveinput{}} }
\newcommand{\intenseveif}[2][]{ \mu_{\detevelabel[#1]|\eveinput{}={#2}} }
\newcommand{\intenseveifvac}[1][]{ \intenseveif[#1]{0} }
\newcommand{\intenseveifcoh}[1][]{ \intenseveif[#1]{\alpha} }

\newcommand{\pnceve}[1][]{ p_{\overline{\detevelabel[#1]}|\eveinput{}} }
\newcommand{\pnceveif}[2][]{ p_{\overline{\detevelabel[#1]}|\eveinput{}={#2}} }
\newcommand{\pnceveifvac}[1][]{ \pnceveif[#1]{0} }
\newcommand{\pnceveifcoh}[1][]{ \pnceveif[#1]{\alpha} }

\newcommand{\gain}{ G }
\newcommand{\qber}{ \textrm{QBER} }
\newcommand{\visibility}[1]{ V_{#1} }
\newcommand{\visave}{ \visibility{\textrm{ave}} }
\newcommand{\maxqber}{ \qber^\textrm{th} }
\newcommand{\minvis}{ \visibility{}^\textrm{th} }

\newcommand{\aveofevents}[1]{ N_{#1} }
\newcommand{\faveofevents}[1]{ n_{#1} }

\newcommand{\plastev}[1]{ p_{#1}^\textrm{last} }
\newcommand{\pedgeev}[1]{ p_{#1}^\textrm{edge} }

\newcommand{\eventsignal}{ \textrm{sig} }

\newcommand{\aveofsigs}{ \aveofevents{\eventsignal} }

\newcommand{\eventclick}{ \textrm{clk} }

\newcommand{\aveofclks}{ \aveofevents{\eventclick} }
\newcommand{\faveofclks}{ \faveofevents{\eventclick} }
\newcommand{\pclickifvac}{ p_{\eventclick|\bob_\textrm{v}} }

\newcommand{\eventdataclk}{ \textrm{key} }

\newcommand{\aveofdataclks}{ \aveofevents{\eventdataclk} }
\newcommand{\faveofdataclks}{ \faveofevents{\eventdataclk} }

\newcommand{\eventerror}{ \textrm{err} }

\newcommand{\aveoferrors}{ \aveofevents{\eventerror} }
\newcommand{\faveoferrors}{ \faveofevents{\eventerror} }
\newcommand{\perrorifvac}{ p_{\eventerror|\bob_\textrm{v}} }

\newcommand{\eventvisclk}[2]{ \detectormonit{#1},{#2} }

\newcommand{\aveofvisclks}[2]{ \aveofevents{\eventvisclk{#1}{#2}} }
\newcommand{\faveofvisclks}[2]{ \faveofevents{\eventvisclk{#1}{#2}} }

\newcommand{\sequence}[1]{``{#1}"}
\newcommand{\selemone}{s_1}
\newcommand{\selemtwo}{s_2}
\newcommand{\svector}{s}
\newcommand{\selems}{\selemtwo\selemone}

\newcommand{\pblocklen}{ p_\textrm{cb} }
\newcommand{\maxblocklen}{{ M_\textrm{max} }}

\newcommand{\palice}[1]{ p_{\alice_{#1}} }

\newcommand{\presifali}[2]{ p_{\eve_{#1}|\alice_{#2}} }
\newcommand{\paliifres}[2]{ p_{\alice_{#1}|\eve_{#2}} }
\newcommand{\presifalivis}[2]{ \presifali{#1}{\selemone}\presifali{#2}{\selemtwo} }

\newcommand{\pconc}{ p_{\eve_c} }
\newcommand{\presifconc}[1]{ p_{\eve_{#1}|\eve_c} }
\newcommand{\paliifconc}[1]{ p_{\alice_{#1}|\eve_c} }

\newcommand{\palidata}{ \palice{\textrm{key}} }
\newcommand{\pdataifres}[1]{ \paliifres{\textrm{key}}{#1} }
\newcommand{\pdataifconc}{ \paliifconc{\textrm{key}} }
\newcommand{\perrifbob}[1]{ p_{\eventerror|\bob_{#1}} }
\newcommand{\perrifbobnvac}{ p_{\eventerror|\bob_c} }
\newcommand{\presifalitoconc}[2]{ p_{\eve_{#1}|\alice_{#2},\eve_c} }
\newcommand{\pcohclick}[1]{ p_{\eventvisclk{#1}{\selemtwo\selemone}|\bob_c} }

\newcommand{\constrecursion}{ R }

\newcommand{\trackchanges}{}
\newcommand{\tracked}[1]{{\trackchanges#1}}

\begin{document}

\preprint{APS/123-QED}

\title{Hacking coherent-one-way quantum key distribution with present-day technology}

\author{Javier  \surname{Rey-Domínguez}}
\email{j.reydominguez@leeds.ac.uk}
\affiliation{School of Electronic and Electrical Engineering, Pollard Institute, University of Leeds, Leeds LS2 9JT, United Kingdom}
\affiliation{Escuela de Ingeniería de Telecomunicación, Department of Signal Theory and Communications, University of Vigo, Vigo E-36310, Spain}

\author{Álvaro \surname{Navarrete}}
\affiliation{Vigo Quantum Communication Center, University of Vigo, Vigo E-36310, Spain}
\affiliation{Escuela de Ingeniería de Telecomunicación, Department of Signal Theory and Communications, University of Vigo, Vigo E-36310, Spain}
\affiliation{AtlanTTic Research Center, University of Vigo, Vigo E-36310, Spain}

\author{Peter \surname{van Loock}}
\affiliation{Johannes-Gutenberg University of Mainz, Institute of Physics, Staudingerweg 7, 55128 Mainz, Germany}

\author{Marcos  \surname{Curty}}
\affiliation{Vigo Quantum Communication Center, University of Vigo, Vigo E-36310, Spain}
\affiliation{Escuela de Ingeniería de Telecomunicación, Department of Signal Theory and Communications, University of Vigo, Vigo E-36310, Spain}
\affiliation{AtlanTTic Research Center, University of Vigo, Vigo E-36310, Spain}

\date{\today}

\begin{abstract}
Recent results have shown that the secret-key rate of coherent-one-way (COW) quantum key distribution (QKD) scales quadratically with the system’s
transmittance, thus rendering this protocol unsuitable for long-distance
transmission.
This was proven by using a so-called zero-error attack, which relies
on an unambiguous state discrimination (USD) measurement.
This type of attack allows the eavesdropper to learn the whole secret key without introducing any error.
Here, we investigate the feasibility and effectiveness of zero-error attacks against COW QKD with present-day technology.
For this, we introduce two practical USD receivers that can be realised with linear passive optical elements, phase-space displacement operations and threshold single-photon detectors.
The first receiver is optimal with respect to its success probability, while the second one can impose stronger restrictions on the protocol’s performance with faulty eavesdropping equipment.
Our findings suggest that zero-error attacks could break the security of COW QKD even assuming realistic experimental conditions.
\end{abstract}

\keywords{quantum key distribution, quantum hacking, coherent-one-way}

\maketitle


\section{Introduction}

Quantum key distribution (QKD)~\cite{Lo_Secure:2014,Xu_Secure:2020,Pirandola_Advances:2020} has emerged as a cornerstone of quantum cryptography, enabling two remote parties, commonly referred to as Alice and Bob, to share an information-theoretically secure cryptographic key.
While QKD networks are currently being deployed worldwide~\cite{Stucki_Long:2011,Sasaki_Field:2011,Qiu_Quantum:2014,Chen_Integrated:2021}, QKD still faces certain inherent limitations such as channel loss, which fundamentally restricts the secret-key rate in point-to-point configurations~\cite{Takeoka_Fundamental:2014,Pirandola_Fundamental:2017}, as well as device imperfections that jeopardize the security of practical implementations~\cite{Xu_Secure:2020,Jain_Attacks:2016,Marquardt_Implementation:2023}.

Various strategies have been proposed to mitigate these limitations and improve the security, practicality, and performance of QKD systems, including \textit{e.g.} decoy-state QKD~\cite{Hwang_Quantum:2003,Lo_DecoyState:2005,Wang_Beating:2005}, measurement-device-independent QKD~\cite{Lo_MDI:2012, Comandar_WOVulnerabilities:2016, Woodward_Gigahertz:2021, Cao_LongDistance:2020, Wei_HighSpeed:2020, Yin_MDIQKD:2016}, twin-field QKD~\cite{Lucamarini_Overcoming:2018,Curty_Simple:2019,Ma_Phase:2018,Lin_Simple:2018,Wang_Twin:2018}, and distributed-phase-reference (DPR) QKD.
Among the latter protocols, coherent-one-way (COW) QKD~\cite{Gisin_Towards:2004,Stucki_FastSimple:2005,Stucki_HighRate:2009,Korzh_Provably:2015} has attracted great attention in recent years for its simplicity and its promise to overcome the photon-number-splitting (PNS) attack~\cite{Huttner_Quantum:1995,Brassard_Limitations:2000}, thus achieving long transmission distances. 
Indeed, commercial systems implementing the COW protocol have been developed~\cite{ID_Quantique} and experimental demonstrations have achieved distances of over 300~km~\cite{Korzh_Provably:2015}.
However, it is important to note that security analyses of COW-QKD that allow for long-distance communications ---\textit{i.e.}, that provide lower bounds on the secret-key rate that scale linearly with the channel transmittance $\eta$--- have been established solely against a restricted class of attacks termed collective attacks~\cite{Korzh_Provably:2015}. This contrasts with known lower bounds of the order of $O(\eta^2)$ against general attacks~\cite{Moroder_SecurityDPR:2012}, a key-rate scaling that has not been improved in recent variants of COW-QKD that disregard its characteristic inter-round interference~\cite{Lavie_Improved:2022,Gao2_Simple:2022,Li_Finite:2024}.

Crucially, González-Payo \textit{et al.}~\cite{Gonzalez_Bounds:2020} showed very recently that indeed the key rate of COW-QKD scales at most quadratically with the system's transmittance, rendering all long-distance demonstrations of this scheme performed so far insecure against general attacks.
This was achieved using a class of intercept-and-resend attacks known as sequential attacks~\cite{Waks_SecurityDPS:2006,Curty_Sequential:2007,Tsurumaru_Sequential:2007,Curty_Effect:2008}.
Intercept-and-resend attacks effectively transform the quantum channel into an entanglement-breaking channel, thus preventing the possibility of secret-key generation~\cite{Curty_Entanglement:2004}.
Trényi \textit{et al.}~\cite{Trenyi_Attack:2021} further refined this strategy by introducing a sequential attack that does not introduce errors in the system.
Notably, this latter so-called zero-error attack ---which is based on the use of unambiguous state discrimination (USD) measurements~\cite{Chefles_USD:1998,Chefles_Optimum:1998}--- is essentially optimal, in the sense that no other zero-error attack~\cite{Branciard_ZeroError:2006} can further limit the maximum achievable distance of COW-QKD.

Importantly, the works in \cite{Gonzalez_Bounds:2020, Trenyi_Attack:2021} assume an idealised eavesdropper (Eve) with technological capabilities only limited by quantum mechanics.
Indeed, this is the standard scenario considered when proving the security of QKD.
However, this could be overconservative in certain cases, as the technology required by Eve to implement her attack might not be available in the mid-term future.
For instance, the noisy-storage model \cite{Wehner_Noisy:2008,Konig_Unconditional:2012,Damgard_Tight:2007,Wehner_TwoParty:2010} considers the physical assumption that Eve does not have a large reliable quantum memory.

Similarly, in this work we assume that Eve is restricted to use present-day technology and cannot perform perfect quantum operations but employs faulty devices.
In this framework, we investigate the practical feasibility of the optimal zero-error attack against COW-QKD proposed in \cite{Trenyi_Attack:2021}.
For this, we introduce two USD receivers that only require off-the-shelf linear passive optical elements, phase-space displacement operations and threshold single-photon detectors (SPDs).
Remarkably, the first receiver corresponds to an optimal USD measurement, in the sense that it maximizes the probability of obtaining a conclusive measurement result when distinguishing Alice’s signals, but it can only
discriminate data signals.
The second USD receiver has a lower success probability but can discriminate both data and decoy signals.
This latter condition translates into stringent restrictions on the performance of COW QKD with flawed eavesdropping equipment.
For both receivers, we derive analytical expressions for the expected values of the key metrics that characterize the COW protocol as a function of the parameters that describe the noise and inefficiencies of Eve’s apparatuses.
In doing so, we provide a comprehensive framework for evaluating the security of COW-QKD in realistic scenarios.
We find that the most critical experimental parameter for the success of Eve’s attack seems to be the quality of interference between Alice’s weak coherent pulses and her strong light during the displacement operation.
Importantly, our results suggest that zero-error attacks are not only a great threat against COW-QKD, but they could break its security with present-day technology.

The paper is structured as follows.
In \cref{sec:COW} we introduce the COW-QKD protocol.
Next, in \cref{sec:zero_err_atatcks}, we present the zero-error attack studied in \cite{Trenyi_Attack:2021}.
Then, in \cref{sec:implementation} we introduce the first USD receiver, which is able to implement the optimal USD measurement considered in \cite{Trenyi_Attack:2021}.
Besides, in this section we provide a model to incorporate its most relevant imperfections in a practical setting.
Next, in \cref{sec:metrics} we derive analytical expressions for the expected values of the relevant metrics required to evaluate the security and performance of COW QKD as a function of Eve’s faulty equipment, and we investigate the feasibility of the zero-error attack in \cite{Trenyi_Attack:2021} with present-day technology in \cref{sec:results}.
Finally, in \cref{sec:conclusions} we present our conclusions.
The paper also includes several Appendices with additional calculations, which includes the analysis associated to a second USD receiver.

\section{Coherent-One-Way QKD}\label{sec:COW}

\begin{figure}
  \includegraphics[width=1\columnwidth]{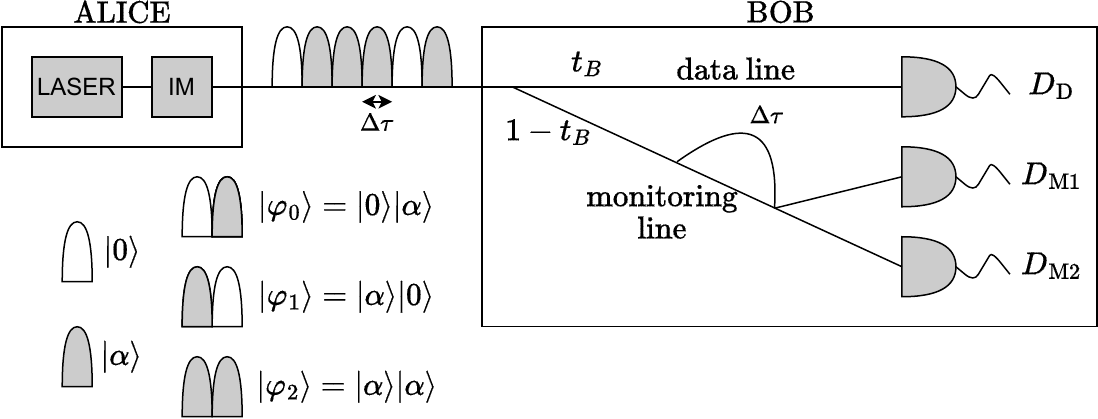}
  \caption{
  Schematic illustration of the original COW system \cite{Gisin_Towards:2004, Stucki_FastSimple:2005}.
  Alice sends Bob a random sequence of signals $\ket{\varphi_i}$, with $i\in\{0,1,2\}$. In the figure, we represent coherent (vacuum) states with a grey (white) pulse. Temporal sequences of states or signals are represented from right (earlier time) to left (latter time). Bob uses a beamsplitter with transmittance $t_B$ to distribute the received signals between the data line and the monitoring line.
  The former is used for key generation, while the latter implements a Mach-Zehnder (MZ) interferometer that measures the coherence between adjacent coherent pulses.
  This is used to detect eavesdropping.
  IM refers to the intensity modulator; $\Delta \tau$ is the time delay between consecutive pulses; $\detbdata$, $\detbmonit[1]$, and $\detbmonit[2]$ are Bob's SPDs.
  }
  \label{fig:cow_scheme}
\end{figure}

The setup for the original COW system~\cite{Gisin_Towards:2004, Stucki_FastSimple:2005} is shown in \cref{fig:cow_scheme}.
In each round, Alice transmits a signal $\cowsig{i}$ to Bob with probability $\palice{i}$, where $i\in\{0,1,2\}$.
These signals are composed of two optical pulses that could either be in a vacuum state $\ket{0}$ or in a coherent state $\ket{\alpha}$, where $\alpha>0$.
Specifically, the data signals $\cowsig{0}=\ket{0}\!\ket{\alpha}$ and $\cowsig{1}=\ket{\alpha}\!\ket{0}$ correspond to the bit values 0 and 1, respectively, and are generated with an equal a priori probability $\palice{0}=\palice{1}=(1-f)/2$, whereas the decoy signal $\cowsig{2}=\ket{\alpha}\!\ket{\alpha}$ is prepared with probability $\palice{2}=f$.
Here, temporal sequences of states or signals are represented from right (earlier time) to left (later time).

At Bob's side, an asymmetric beamsplitter with transmittance $t_B$ distributes the incoming signals between the data line and the monitoring line.
The former consists of an SPD $\detbdata$ and is used for raw key generation.
Specifically, Bob assigns the bit value 0 (1) to a round if $\detbdata$ clicks in the first (second) time slot of that round, and a random bit is assigned in the case of a double click.
On the other hand, the monitoring line consists of a Mach-Zehnder interferometer followed by two SPDs $\detbmonit[1]$ and $\detbmonit[2]$, for constructive and destructive interference, respectively.
This line monitors the coherence between adjacent pulses.
In particular, the interference is such that two consecutive coherent states $\ket{\alpha}$ cannot trigger $\detbmonit[2]$, but only $\detbmonit[1]$ (see \cref{fig:cow_scheme}).

Once the quantum communication phase of the protocol ends, Bob publicly announces in which rounds he observed at least one detection click at $\detbdata$.
Then, Alice announces in which of these rounds she prepared a data signal.
The bits assigned to this set constitute the sifted key.

Three parameters are specially relevant in COW-QKD.
The gain, $\gain$, which is the probability that a signal sent by Alice produces at least one detection click in Bob's data line; the quantum bit error rate ($\qber$) of Alice and Bob's sifted keys; and the visibilities $\visibility{\svector}$, which quantify the coherence between adjacent coherent pulses, and are computed from the click probabilities in the monitoring line.
Specifically, these visibilities are defined as
\begin{equation}\label{eq:visibilities}
    \visibility{\svector}:= \frac{ p_{\detbmonitlabel[1]|\svector} - p_{\detbmonitlabel[2]|\svector} }{ p_{\detbmonitlabel[1]|\svector} + p_{\detbmonitlabel[2]|\svector} },
\end{equation}
where $\svector \in \{ 2, 01, 02, 21, 22 \}$ represents a sequence of COW signals that contains two adjacent coherent states $\ket{\alpha}$, and $p_{\detbmonitlabel[i]|\svector}$ is the conditional probability that detector $\detbmonit[i]$ clicks when such two coherent pulses interfere, given that Alice prepared the sequence $\svector$.
For example, $\visibility{2}$ characterises the visibility between the two coherent pulses contained in a decoy signal $\cowsig{2}$, whereas $\visibility{02}$ characterises the visibility between the second optical pulse of $\cowsig{2}$ and the first one of $\cowsig{0}$, both of them in the state $\ket{\alpha}$.
The other visibilities are interpreted in a similar way.
Finally, it is convenient to consider the average visibility, which is given by \cite{Korzh_Provably:2015, Gonzalez_Bounds:2020}
\begin{equation}\label{eq:visave}
    \visave :=
    \frac{ p_{\detbmonitlabel[1]} - p_{\detbmonitlabel[2]} }{ p_{\detbmonitlabel[1]} + p_{\detbmonitlabel[2]} },
\end{equation}
where $p_{\detbmonitlabel[i]}=\sum_{\svector}p_{\svector}p_{\detbmonitlabel[i]|\svector}$, being $p_{\svector}$ the probability that Alice prepares the sequence $\svector$.

\section{Zero-error attacks against COW-QKD}\label{sec:zero_err_atatcks}

In a zero-error attack Eve intercepts all of Alice's signals and performs a USD measurement on each of them \cite{Gonzalez_Bounds:2020, Trenyi_Attack:2021}.
We denote by $\presifali{j}{i}$ the probability that Eve obtains the result $\eve_j$, given that Alice emits the signal $\cowsig{i}$.
Here, $\eve_0$, $\eve_1$ and $\eve_2$ identify the signals $\cowsig{0}$, $\cowsig{1}$ and $\cowsig{2}$, respectively, and $\eve_3$ represents an inconclusive outcome.
Obviously, in the ideal scenario in which the USD measurement is implemented perfectly, we have by definition that $\presifali{j}{i} = 0 \ \forall i \neq j$, with $i,j < 3$.

Next, Eve groups the measured outcomes into blocks for processing them before she sends Bob a regenerated sequence of signals.
Precisely, a block of $(k+1)$ signals corresponds to $k\in\set{0,1,\dots,\maxblocklen}$ consecutive conclusive measurement outcomes, followed by an inconclusive measurement result.
For example, when $k=0$ the block corresponds to one inconclusive measurement outcome, and Eve sends Bob two vacuum pulses, $\vacsig= \ket{0}\!\ket{0}$.
If $k=1$, the block has one conclusive measurement outcome, say $\cowsig{i}$, followed by an inconclusive one for which Eve sends Bob the vacuum signal $\vacsig$.
For all cases where $1<k<\maxblocklen$, the interpretation is similar.
Finally, if Eve obtains $\maxblocklen$ consecutive conclusive measurement outcomes, she ignores the next signal from Alice, and simply treats it as an inconclusive result. 
The parameter $\maxblocklen$ allows Eve to cap the block length, thereby limiting the maximum delay she introduces in the channel.
Throughout this paper, we shall use the term \textit{block} to denote the $k$ conclusively measured signals together with the inconclusive measurement outcome, and the term \textit{conclusive-block} when ignoring the latter.

For each conclusive-block, Eve searches for the first and last instances of a vacuum pulse within the block.
She then resends to Bob all the optical pulses situated between these two, exactly as she identified them ---\textit{i.e.}, if the measurement result with respect to a particular signal is $\eve_j$, she resends $\cowsig{j}$--- but substitutes the coherent pulses $\ket{\alpha}$ by $\ket{\gamma}$, with $|\gamma|^2 \gg |\alpha|^2$, to increase the detection probability at Bob's side.
The remaining pulses within the block that are outside this interval are resent to Bob as vacuum pulses.
As already mentioned, the last signal of a block, which corresponds to an inconclusive measurement outcome, is resent as $\vacsig$.
For a more detailed description of the attack, we refer the reader to \cite{Trenyi_Attack:2021}.

Notably, if Eve's equipment is flawless ---as considered in~\cite{Trenyi_Attack:2021}--- the attack described above introduces no errors on Bob's side, while it reveals Eve full information about the key.
This is because Eve's USD measurement ensures that she never misidentifies Alice's signals, while the block processing strategy takes advantage of the fact that vacuum pulses do not introduce errors in the monitoring line.

In particular, \cite{Trenyi_Attack:2021} showed that, whenever $f/(1-f) \leq 2 e^{-|\alpha|^2}$, a regime typically satisfied by practical implementations of COW-QKD, Eve's optimal USD measurement (\textit{i.e.}, the one that maximizes her probability of a conclusive result) satisfies $\presifali{0}{0}=\presifali{1}{1}=1-e^{-|\alpha|^2}$ and  $\presifali{0}{1} = \presifali{1}{0} = \presifali{2}{i}=0 \;\forall i \in \set{0,1,2}$.

\section{Implementation of zero-error attacks}\label{sec:implementation}

\begin{figure}
  \includegraphics[width=1\columnwidth]{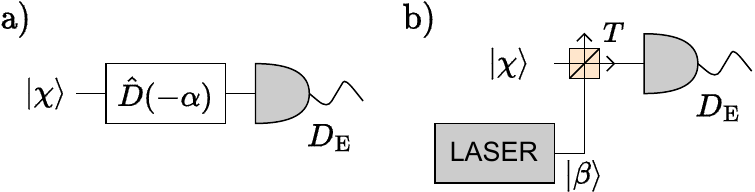}
  \caption{Scheme to implement Eve's optimal USD measurement.
  a) Ideal theoretical setup: Eve applies a displacement operator $\hat{D}(-\alpha)$ on each pulse $\ket{\eveinput}$ of the signal, and subsequently measures the outcome pulses with a threshold detector $\deteve$.
  b) Experimental implementation of the ideal setup: The displacement operation can be approximated in practice with an asymmetric beamsplitter of transmittance $T \approx 1$, which has at one of its input ports a laser source that emits coherent pulses with amplitude $\beta$.}
  \label{fig:scheme_optimal}
\end{figure}

In this section, we now introduce a linear optics circuit to implement Eve's optimal USD measurement for the zero-error attack described above.
It is illustrated in \cref{fig:scheme_optimal}a.
The input state $\ket{\eveinput}$, with $\eveinput\in\{0,\alpha\}$, corresponds to each of the two optical pulses sent by Alice within a signal.
That is, for each signal, Eve uses the same scheme in \cref{fig:scheme_optimal}a twice, once per pulse.
Precisely, each pulse is displaced according to the transformation $\ket{\eveinput}\to\hat D(-\alpha)\ket{\eveinput}=\ket{\eveinput-\alpha}$, where $\hat D(x):=e^{x\hat{a}^{\dagger}-x^*\hat{a}}$ is the displacement operator, and $\hat{a}^{\dagger}$ and $\hat{a}$ are the creation and annihilation operators, respectively.
Finally, the resulting signal is measured with a SPD.

If $\deteve$ clicks only in the first (second) time slot, the signal is identified as $\cowsig{1}$ ($\cowsig{0}$); in all other cases, the result is inconclusive.
This is so because after the optical displacement, the data signals $\cowsig{0}$ and $\cowsig{1}$ are transformed into $\ket{-\alpha}\ket{0}$ and $\ket{0}\ket{-\alpha}$, respectively, while the decoy signal is turned into a vacuum signal $\vacsig$. Consequently, a single click uniquely identifies a data signal, which occurs with probability $1-e^{-|\alpha|^2}$, thus matching the optimal probability obtained in \cite{Trenyi_Attack:2021}.
Indeed, this measurement is unable to identify decoy signals, effectively removing $\eve_2$ from the POVM set and making $\presifali{2}{i}=0 \ \forall i$.

In practice, it is well-known that a displacement operation can be approximated with a highly asymmetric beamsplitter of transmittance $T \approx 1$ \cite{Paris_Displacement:1996}, which has at one of its input ports a coherent state $\ket{\beta}$, as shown in \cref{fig:scheme_optimal}b.
This scheme transforms $\ket{\eveinput}$ 
into $\ket{\sqrt{T}\eveinput+\sqrt{1-T}\beta}$, so the displacement $\hat D(-\alpha)$ can be approximated by setting
\begin{equation}\label{eq:interfering_usd1}
    \beta= -\sqrt{\frac{T}{1-T}} \alpha.
\end{equation}

Indeed, with this choice, the states $\ket{0}$ and $\ket{\alpha}$ are transformed, respectively, into $\ket{-\sqrt{T}\alpha}$ and $\ket{0}$.
This means that the detection probability at $\deteve$ decreases slightly when Alice transmits $\ket{0}$ (when compared to the case where Eve can use an ideal displacement).
However, it remains zero when Alice sends $\ket{\alpha}$.
That is, the approximated displacement does not introduce errors.

\subsection{Effect of device imperfections}\label{sec:dev_imperfect}

Now, we investigate the performance of the setup above in a realistic setting, in which we accommodate the most relevant device imperfections.
For this, we allow the optical phase of the incoming pulses $\ket{\eveinput}$ to the beamsplitter in \cref{fig:scheme_optimal}b to be slightly shifted with respect to the laser pulses $\ket{\beta}$.
We denote such phase shift by $\phi$.
Also, we allow for an imperfect mode overlap between the two interfering pulses $\ket{\eveinput e^{i\phi}}$ and $\ket{\beta}$ at the beamsplitter.
To characterize this effect we use the model introduced in \cite{Laiho_Probing:2009}, which defines two parameters, $t_1$ and $t_2$, to quantify, respectively, the fraction of each of Alice's and Eve's input pulses that is properly mode matched at the beamsplitter (see \cref{sec:mode_mismatch}).
In addition, we allow for a non-ideal efficiency $\effeve$ and a dark-count probability $\darkcount{\detevelabel}$ in Eve's SPD $\deteve$.
Finally, we also account for small intensity fluctuations.
For this, we consider a simple model in which the amplitude of $\ket{\beta}$ is slightly deviated from its ideal value (see \cref{eq:interfering_usd1}). In particular, we compute its amplitude, which we call now $\sigma$ to distinguish it from the ideal case, as
\begin{equation}\label{eq:usd1_beta_err}
    \sigma =
    \sqrt{1 + \amperror} \beta,
\end{equation}
where the parameter $\amperror \!\in\! [-1,\infty)$ characterizes the deviation between the ideal and actual intensity.

\begin{figure}
  \includegraphics[width=1\columnwidth]{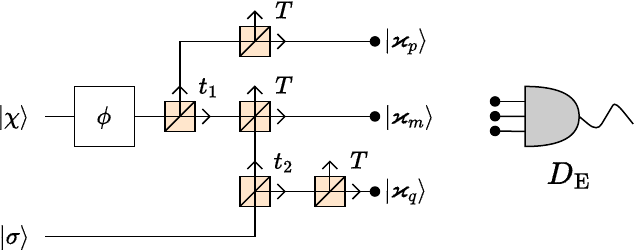}
  \caption{Theoretical model of the setup illustrated in \cref{fig:scheme_optimal}b, now incorporating the effect of the main experimental imperfections. $\ket{\eveinput}$ represents the state of Alice's transmitted pulse, and $\phi$ represents a phase shift. The beamsplitter transmittances $t_1$ and $t_2$ quantify the mode overlap between Alice's pulse and a coherent state $\ket{\sigma}$, whose intensity deviates slightly from that of $\ket{\beta}$, at the beamsplitter with transmittance $T$ (see ~\cref{fig:scheme_optimal}b).
  The three outputs of the circuit represent different optical modes in coherent states $\ket{\varkappa_x}$, with $x \in \{m, p, q\}$.
  The total intensity received in $\deteve$ is calculated as the sum of the intensities in each of these modes.}
  \label{fig:model_optimal}
\end{figure}

The schematic of the model that incorporates these imperfections is illustrated in \cref{fig:model_optimal}.
First, the incoming pulse $\ket{\eveinput}$ is phase shifted by $\phi$.
Then, $\ket{\eveinput e^{i\phi}}$ and $\ket{\beta}$ are split into two modes each, according to the quantities $t_1$ and $t_2$.
Two of these optical modes contain the fraction of the input pulses that properly interfere at the beamsplitter with transmittance $T$.
We use the label ``$m$" to denote the optical mode associated with the output port of this latter beamsplitter that is connected to Eve's detector.
The remaining output modes, namely those labelled ``$p$" and ``$q$" in \cref{fig:model_optimal}, go through the beamsplitter without interfering.
The total optical intensity at $\deteve$ given that Alice sent the state $\ket{\eveinput}$, which we shall denote as $\intenseve$, is thus the sum of the intensities from the three output modes.
That is, $\intenseve = |\varkappa_m|^2 + |\varkappa_p|^2 + |\varkappa_q|^2$, where $\ket{\varkappa_x}$ denotes a coherent state in the output mode $x$, with $x \in \{m, p, q\}$.
We find, therefore, that
\begin{equation}\label{eq:intensity_eve}\begin{split}
    \intenseve =\:& T \Big[
    |\eveinput|^2
    + |\alpha|^2 \left( 1 + \amperror \right)
    \\
    -& 2 \sqrt{t_1t_2 \left( 1 + \amperror \right)} \textrm{Re}\{\eveinput \alpha^* e^{i\phi} \}
    \Big].
\end{split}\end{equation}

Let $\pnceveifvac$ ($\pnceveifcoh$) denote the probability of a no-click event in $\deteve$, given that $\eveinput=0$ ($\eveinput=\alpha$) in that time slot.
These probabilities can be computed as $\pnceve = (1-\darkcount{\detevelabel}) \exp\{ -\effeve \intenseve \}$.
That is,
\begin{equation}\begin{split}
    \intenseveifvac =\:&
    T|\alpha|^2 \left( 1 + \amperror \right),
    \\
    \intenseveifcoh =\:&
    T|\alpha|^2 \left[
    2 + \amperror
    - 2\sqrt{t_1t_2\left( 1 + \amperror \right)} \cos\phi
    \right].
\end{split}\end{equation}
This means, in particular, that the probabilities $\presifali{j}{i}$ can be expressed as follows:
\begin{equation}\label{eq:usdm1_probs}\begin{split}
    &\presifali{0}{0} =
    \presifali{1}{1} =
    \pnceveifcoh \left( 1 - \pnceveifvac \right),
    \\
    &\presifali{0}{1} =
    \presifali{1}{0} =
    \pnceveifvac \left( 1 - \pnceveifcoh \right),
    \\
    &\presifali{0}{2} =
    \presifali{1}{2} =
    \pnceveifcoh \left( 1 - \pnceveifcoh \right),
    \\
    &\presifali{2}{0} = \presifali{2}{1} = \presifali{2}{2} = 0,
    \\
    &\presifali{3}{i} = 1 - \left( \presifali{0}{i} + \presifali{1}{i} \right) \;\; \textrm{for } i \in \{0,1,2\}.
\end{split}\end{equation}

\section{Performance evaluation}\label{sec:metrics}

To evaluate the performance of Eve's zero-error attack with current technology, here we derive analytical expressions for the expected values of the three metrics defined in~\Cref{sec:COW} ---namely the gain, the QBER, and the different visibilities $\visibility{\svector}$--- as a function of the parameters that characterize Eve's imperfect operation, as well as the parameters of the protocol.

In the calculations below, we shall consider that $\gamma$ is sufficiently large to ensure that the signals Eve sends Bob always trigger his detectors unless she sends him a vacuum state, in which case Bob only records a click if a dark count occurs.
The dark-count probabilities at Bob's detectors $\detbdata$, $\detbmonit[1]$, and $\detbmonit[2]$ are denoted as $\darkcount{\detectordata}$, $\darkcount{\detectormonit{1}}$, and $\darkcount{\detectormonit{2}}$, respectively.

First, we derive the expected gain $G$.
Then, we present the expressions to compute the QBER and the visibilities $\visibility{\svector}$.
The full derivation of these latter parameters can be found in \cref{sec:SM_metrics}.

\subsection{Gain}\label{sec:metrics_gain}
The gain is defined as the probability that Bob observes at least one click in $D_{\detectordata}$ in a round.
This quantity can be written as $\gain= \aveofclks/\aveofsigs$, where $\aveofclks$ is the average number of signals within a block that produce a click ---single or double--- at Bob's side, and $\aveofsigs$ is the average number of signals within a block.

Let $\pconc$ be the probability that Eve's USD measurement is conclusive, computed as
\begin{equation}\label{eq:pconc}
    \pconc =
    \sum_{i,j=0}^2 \palice{j}\presifali{i}{j}.
\end{equation}
Also, let $\pblocklen(k)$ be the probability that Eve processes a conclusive-block of length $k$, which is given by \cite{Trenyi_Attack:2021}
\begin{equation}
    \pblocklen(k) =
    \begin{cases}
        \pconc^k (1-\pconc)
        &\textrm{when } 0 \leq k < \maxblocklen,
        \vspace{0.15cm}
        \\
        \pconc^\maxblocklen
        &\textrm{when } k=\maxblocklen,
        \vspace{0.15cm}
        \\
        0
        &\textrm{otherwise}.
    \end{cases}
\end{equation}
Then, we have that $\aveofsigs= \sum_{k=0}^\maxblocklen \pblocklen(k) (k+1)$, and therefore \cite{Trenyi_Attack:2021}
\begin{equation}\begin{split}
    &\aveofsigs =
    \frac{1 - \pconc^{\maxblocklen+1}}{1-\pconc}.
\end{split}\end{equation}

The average $\aveofclks$ admits a similar decomposition in terms of the length of a block. In particular,
\begin{equation}\label{eq:avefromfave_clks}\begin{split}
    &\aveofclks =
    \sum_{k=0}^\maxblocklen \pblocklen(k) \left[
    \faveofclks(k) + \plastev{\eventclick}(k) \right],
\end{split}\end{equation}
where $\faveofclks(k)$ is the average number of signals within a conclusive-block of length $k$ in which Bob observes at least one click in $D_{\detectordata}$, and $\plastev{\eventclick}(k)$ is the probability that a click occurs in the last vacuum signal $\vacsig$ of the block, corresponding to the inconclusive measurement outcome that happened after a conclusive-block of length $k$.
If we define $\pclickifvac$ as the probability that Bob observes at least one click in a round where he receives a vacuum signal from Eve, then it is clear that
\begin{equation}\label{eq:gain_last}
    \plastev{\eventclick}(k) =
    \pclickifvac =
    1 - \left( 1 - \darkcount{\detectordata} \right)^2,
\end{equation}
for $k \in [0,\maxblocklen]$.

Now, let $\presifconc{j}$ be the probability that Eve obtains the outcome $\eve_{j}$, given that her measurement was conclusive. Its value is given by
\begin{equation}\label{eq:presifconc}
    \presifconc{j}=
    \frac{1}{\pconc}
    \sum_{i=0}^2 \palice{i} \presifali{j}{i}.
\end{equation}
Then, one can express
\begin{equation}\label{eq:n_click_k}
    \faveofclks(k) = \sum_{i=0}^2 \presifconc{i} \faveofclks(k|i),
\end{equation}
where  $\faveofclks(k|i)$ is defined as the average number of signals where a click occurs within a conclusive-block of length $k$, given that the first signal of the block was identified by Eve as $\cowsig{i}$.
Importantly, we note that the quantities $\faveofclks(k|i)$ admit recursive formulations \cite{Trenyi_Attack:2021}, which one can solve to compute $\faveofclks(k)$.

To illustrate this, let us focus on $\faveofclks(k|0)$, \textit{i.e.}, the case where the first signal of the block is identified by Eve as $\cowsig{0}$.
According to her block-processing strategy, Eve translates this first signal into vacuum, since the first optical pulse of this signal is a coherent state $\ket{\alpha}$.
Consequently, we can disregard it and consider the reduced conclusive-block of length $(k-1)$.
Importantly, according to Eve's processing strategy, this also implies that the first signal in the resulting truncated conclusive-block will be resent exactly as identified by her.
This is so because it is preceded by a vacuum pulse, given by the second optical pulse of $\cowsig{0}$.
Then, it only remains to find the last signal of the conclusive-block that is not resent as vacuum.

There are three possibilities, which depend on the last signal identified by Eve.
If this signal is $\cowsig{0}$, then it directly becomes the last non-vacuum signal of the block, and thus $(k-1)$ non-vacuum signals will arrive at $D_\detectordata$.
If the last signal is $\cowsig{1}$, it is translated into vacuum, and the preceding $(k-2)$ non-vacuum signals will arrive at $D_\detectordata$.
Finally, if the last signal is $\cowsig{2}$, this signal is also translated into vacuum, and, by definition, Bob will observe, on average, $\faveofclks(k-1|0)$ clicks in $D_\detectordata$.
Putting all this together, one can write $\faveofclks(k-1|i)$ as
\begin{equation}\begin{split}
    \faveofclks(k|0) =\:&
    \presifconc{0} \left( k - 1 + \pclickifvac \right)
    \\
    +\:& \presifconc{1} \left( k - 2 + 2 \pclickifvac \right)
    \\
    +\:& \presifconc{2} \left[ \faveofclks(k-1|0) + \pclickifvac \right],
    \\ 
    \faveofclks(k|1) =\:&
    \presifconc{0} k
    \\
    +\:& \presifconc{1} \left( k - 1 + \pclickifvac \right)
    \\
    +\:& \presifconc{2} \left[ \faveofclks(k-1|1) + \pclickifvac \right],
    \\ 
    \faveofclks(k|2) =\:&
    \faveofclks(k-1)
    + \pclickifvac.
\end{split}\end{equation}
Moreover, the starting points for the previous recursions are $\faveofclks(1|0)= \faveofclks(1|1)= \pclickifvac$ and $\faveofclks(0)= 0$. With this, one can solve the recursion and obtain
\begin{equation}\label{eq:gain_faves}\begin{split}
    \faveofclks(k) =\:&
    k
    + k \constrecursion^k \left( 1 - \darkcount{\detectordata} \right)^2
    \\
    -\:& \frac{1+\constrecursion}{1-\constrecursion} \left( 1 - \constrecursion^k \right) \left(1 - \darkcount{\detectordata} \right)^2,
\end{split}\end{equation}
where we have used $\constrecursion$ as a shorthand for the recursion factor,
\begin{equation}\label{eq:constrecursion}
    \constrecursion \equiv \presifconc{2}.
\end{equation}

\subsection{Quantum bit error rate}\label{sec:metrics_qber}
Next, we analyse the $\qber$. 
This quantity can be written as $\qber = p_{\eventerror}/ p_{\eventdataclk}$, where $p_{\eventdataclk}$ is the probability that Bob distills a key bit in a given round, and $p_{\eventerror}$ the probability that he distills an erroneous key bit.
We have that, asymptotically, $p_{\eventdataclk}= \aveofdataclks/ \aveofsigs$ and $p_{\eventerror}= \aveoferrors/ \aveofsigs$, where $\aveofdataclks$ ($\aveoferrors$) is the average number of key bits (erroneous key bits) distilled by Bob from a block sent by Eve.
Therefore, one can rewrite the error rate as \cite{Trenyi_Attack:2021}
\begin{equation}\label{eq:qber_asymptotic}
    \qber=
    \frac{\aveoferrors}{\aveofdataclks}.
\end{equation}

To determine $\aveofdataclks$ and $\aveoferrors$ we decompose them according to the length of the block processed by Eve.
For this, we define $\faveofdataclks(k)$ ($\faveoferrors$(k)) as the average number of key bits (erroneous key bits) distilled from a conclusive-block of length $k$, and $\plastev{\eventdataclk}(k)$ ($\plastev{\eventerror}(k)$) as the probability that Bob distills a key bit (erroneous key bit) from the vacuum signal $\vacsig$ that is sent after a conclusive-block of length $k$, due to an inconclusive result.
Then, we have that
\begin{equation}\label{eq:avefromfave_qber}\begin{split}
    &\aveofdataclks =
    \sum_{k=0}^\maxblocklen \pblocklen(k) \left[ \faveofdataclks(k) + \plastev{\eventdataclk}(k) \right],
    \\
    &\aveoferrors =
    \sum_{k=0}^\maxblocklen \pblocklen(k) \left[ \faveoferrors(k) + \plastev{\eventerror}(k) \right].
\end{split}\end{equation}

All that remains is to calculate the values of $\faveofdataclks(k)$, $\faveoferrors$(k), $\plastev{\eventdataclk}(k)$ and $\plastev{\eventerror}(k)$ that appear in the previous equations. For this, let us define $\perrorifvac$ as the probability that Bob obtains an incorrect bit from a vacuum signal sent by Eve, given that he distills a bit that round.
Its value is given by
\begin{equation}\label{eq:perrifvac}
    \perrorifvac =
    \frac{ \darkcount{\detectordata} \left(2-\darkcount{\detectordata}\right) }{2}.
\end{equation}
Then, it can be shown (see \cref{sec:SM_metrics}) that $\faveofdataclks(k)$ has the form
\begin{equation}\begin{split}\label{eq:n_dataclk}
    \faveofdataclks(k) =\:&
    \frac{\palice{0}}{\pconc} \Big\{
    k \left( 2 - \presifali{3}{0} - \presifali{3}{1} \right)
    \\
    +\:& k \constrecursion^{k-1}  \left( 1 - \darkcount{\detectordata} \right)^2 \left( \presifali{2}{0} + \presifali{2}{1} \right)
    \\
    -\:&
    \frac{1-\constrecursion^k}{1-\constrecursion} \left( 1 - \darkcount{\detectordata} \right)^2
    \big( 2 - \presifali{3}{0} - \presifali{3}{1}
    \\
    +\:& \presifali{2}{0} + \presifali{2}{1} \big)
    \Big\};
\end{split}\end{equation}
the parameter $\plastev{\eventdataclk}$ is given by
\begin{equation}\label{eq:plast_dataclk}\begin{split}
    &\plastev{\eventdataclk}(k)
    \\
    &= \begin{cases}
        \frac{\palice{0} \pclickifvac \left( \presifali{3}{0} + \presifali{3}{1} \right)}
        {1 - \pconc}
        &\textrm{ if } 0 \leq k < \maxblocklen,
        \\
        2 \palice{0} \pclickifvac
        &\textrm{ if } k = \maxblocklen,
    \end{cases}
\end{split}\end{equation}
the parameter $\faveoferrors(k)$ has the form
\begin{equation}\begin{split}\label{eq:n_error}
    \faveoferrors(k) =\:&
    \frac{\palice{0}}{2\pconc} \Big\{
    k \Big[
    \left( \presifali{0}{0} + \presifali{1}{1} \right) \darkcount{\detectordata}
    + \presifali{2}{0}
    \\
    +\:& \presifali{2}{1}
    + \left( \presifali{0}{1} + \presifali{1}{0} \right) \left( 2 - \darkcount{\detectordata} \right)
    \Big]
    \\
    +\:&
    k \constrecursion^{k-1} \left( \presifali{2}{0} + \presifali{2}{1} \right) \left( 1 - \darkcount{\detectordata} \right)^2
    \\
    +\:& \frac{1 - \constrecursion^k}
    {1-\constrecursion}
    \Big[
    \left( \presifali{0}{0} + \presifali{1}{1} \right) \darkcount{\detectordata}
    \\
    -\:&
    \left( \presifali{0}{1} + \presifali{1}{0} \right) \left( 2 - \darkcount{\detectordata} \right)
    \\
    -\:& 2 \!\left( \presifali{2}{0} + \presifali{2}{1} \right)\! \left( 1 - \darkcount{\detectordata} \right)\!
    \Big] \!\left( 1 - \darkcount{\detectordata} \right)\!
    \Big\} ,
\end{split}\end{equation}
and the parameter $\plastev{\eventerror}(k)$ can be expressed as
\begin{equation}\label{eq:plast_error}\begin{split}
    &\plastev{\eventerror}(k)
    \\
    &= \begin{cases}
        \frac{ \palice{0} \perrorifvac \left( \presifali{3}{0} + \presifali{3}{1} \right) }
        {1 - \pconc}
        &\textrm{ if } 0 \leq k < \maxblocklen,
        \\
        2 \palice{0} \perrorifvac
        &\textrm{ if } k = \maxblocklen.
    \end{cases}
\end{split}\end{equation}

\subsection{Visibilities}\label{sec:metrics_vis}
To calculate the resulting visibilities $\visibility{\svector}$, we start by expressing the probabilities that appear in~\cref{eq:visibilities} as $p_{\detectormonit{X}|\svector}= p_{\eventvisclk{X}{\svector}}/p_{\svector}$, with $X \in \{1,2\}$.
The joint probabilities $p_{\eventvisclk{X}{\svector}}$ can be written as $p_{\eventvisclk{X}{\svector}}= \aveofvisclks{X}{\svector}/\aveofsigs$.
Here, $\aveofvisclks{X}{\svector}$ is the average number of times in which the sequence $\svector$ appears within a block and a click is registered in $D_{\detectormonit{X}}$ during the time slot associated with the interference of the two intermediate coherent pulses of the sequence.
This means that $\visibility{\svector}$ can be expressed as
\begin{equation}
    \visibility{\svector} =
    \frac{
    \aveofvisclks{1}{\svector} - \aveofvisclks{2}{\svector}
    }{
    \aveofvisclks{1}{\svector} + \aveofvisclks{2}{\svector}
    }.
\end{equation}

Let us now start with the case $\svector=2$, since this is the only one that considers interference between pulses within the same signal.
By applying analogous reasoning to that used to derive~\cref{eq:avefromfave_clks,eq:avefromfave_qber}, it can be shown that $\aveofvisclks{X}{2}$ can be expressed as
\begin{equation}\label{eq:aveofvis2}
    \aveofvisclks{X}{2}=
    \sum_{k=0}^\maxblocklen \pblocklen(k) \left[ \faveofvisclks{X}{2}(k) + \plastev{\eventvisclk{X}{2}}(k) \right],
\end{equation}
where $\faveofvisclks{X}{2}(k)$ represents the average number of signals within a conclusive block of length $k$ in which Alice sends $\cowsig{2}$ and $\detbmonit[X]$ clicks, and $\plastev{\eventvisclk{X}{2}}(k)$ is the probability that this event occurs in the last signal of the full block.
Precisely, it is shown in \cref{sec:SM_metrics} that these quantities can be written as
\begin{equation}\begin{split}\label{eq:n_M12_M22}
    \faveofvisclks{1}{2}(k) =\:&
    \frac{\palice{2}}{\pconc} \bigg\{
    k \left( 1 - \presifali{3}{2} \right)
    \\
    +\:& k \constrecursion^{k-1} \presifali{2}{2} \left( 1 - \darkcount{\detectormonit{1}} \right)
    \\
    -\:&
    \frac{1 - \constrecursion^k}{1 - \constrecursion} \big( 1 + \presifali{2}{2}
    \\
    -\:& \presifali{3}{2} \big) \left( 1 - \darkcount{\detectormonit{1}}\right)
    \bigg\},
    \\ 
    \faveofvisclks{2}{2}(k) =\:&
    \frac{\palice{2}}{\pconc} \bigg\{
    k \left( \presifali{0}{2} + \presifali{1}{2} \right)
    \\
    -\:& k \constrecursion^{k-1} \presifali{2}{2} \darkcount{\detectormonit{2}}
    \\
    -\:& \frac{1 - \constrecursion^k}{1 - \constrecursion} \Big[
    \left( \presifali{0}{2} + \presifali{1}{2} \right) \left( 1 - \darkcount{\detectormonit{2}} \right)
    \\
    -\:&
    2 \presifali{2}{2} \darkcount{\detectormonit{2}}
    \Big]
    \bigg\},
\end{split}\end{equation}
and
\begin{equation}\label{eq:plast_vis2}
    \plastev{\eventvisclk{X}{2}}(k) =
    \begin{cases}
        \frac{ \palice{2} \presifali{3}{2} \darkcount{\detectormonit{X}} }
        {1 - \pconc}
        & \textrm{if } 0 \leq k < \maxblocklen,
        \vspace{0.3cm}
        \\
        \palice{2}\darkcount{\detectormonit{X}}
        & \textrm{if } k = \maxblocklen,
    \end{cases}
\end{equation}
where $\constrecursion$ is given by \cref{eq:constrecursion}.

We now consider the remaining visibilities, where we denote the two-signal interference sequence as $\svector\equiv \selems$, meaning Alice first prepares $\cowsig{\selemone}$ and subsequently prepares $\cowsig{\selemtwo}$.
Importantly, a subtle nuance must be considered in the definition of the averages $\aveofvisclks{X}{\selemtwo\selemone}$.
Since we are dealing with the interference between adjacent pulses in consecutive rounds, these rounds may belong to different blocks.
If this happens, we will consider that the observed clicks are attributed to the first block. 
This is an arbitrary decision, but it does not impact the final result.

\begin{figure}
  \includegraphics[width=1\columnwidth]{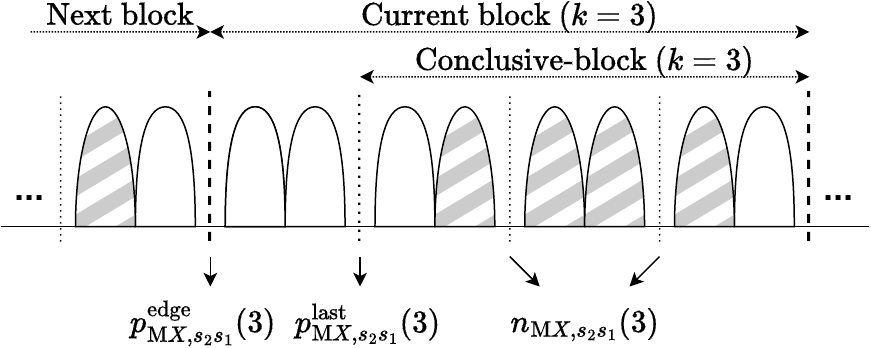}
  \caption{
  Schematic illustration of the different interference events at Bob's side considered within a block, and the quantities involved in the computation of $\aveofvisclks{X}{\selems}$ (see \cref{eq:avefromfave_vis}).
  For simplicity, this figure shows a block with $k=3$.
  White pulses represent vacuum states, $\ket{0}$, sent to Bob by Eve, whereas the stripped pulses may be either $\ket{0}$ or $\ket{\alpha}$.
  Being a valid block, it has two vacuum pulses at the conclusive-block's surrounding borders, as well as a vacuum signal $\vacsig$ at the end of the block.}
  \label{fig:example_visibs}
\end{figure}

Once again, we express the averages under analysis as a decomposition over the block length.
For this, we define $\faveofvisclks{X}{\selemtwo\selemone}(k)$ as the average number of times within a conclusive-block of length $k$ where Alice sends $\svector=\selemtwo\selemone$ and a click is registered in $D_{\detectormonit{X}}$ in the time slot associated with the interference between the two signals.
Those events in which one of the signals of the sequence $\svector$ does not belong to the considered conclusive-block are not accounted in $\faveofvisclks{X}{\selems}(k)$.
Moreover, we denote as $\plastev{\eventvisclk{X}{\selemtwo\selemone}}$ the probability that Alice prepares $\cowsig{\selemone}$ in the last round of a conclusive-block of length $k$ and $\cowsig{\selemtwo}$ in the next round, and Bob observes a click in $D_{\detectormonit{X}}$ in the time slot associated with the interference of these two signals.
Similarly, we define $\pedgeev{\eventvisclk{X}{\selemtwo\selemone}}$ as the probability that Alice prepares $\cowsig{\selemone}$ in the last round of the full block of $k+1$ signals, $\cowsig{\selemtwo}$ in the first round of the next block, and Bob observed a click in $D_{\detectormonit{X}}$ in the time slot associated with the interference of these two signals.
With these definitions, which are illustrated in \cref{fig:example_visibs}, we can write $\aveofvisclks{X}{\selems}$ as
\begin{equation}\label{eq:avefromfave_vis}\begin{split}
    \aveofvisclks{X}{\selemtwo\selemone} =\:&
    \sum_{k=0}^\maxblocklen \pblocklen(k) \Big[
    \faveofvisclks{X}{\selemtwo\selemone}(k)
    \\
    +\:& \plastev{\eventvisclk{X}{\selemtwo\selemone}}(k)
    + \pedgeev{\eventvisclk{X}{\selemtwo\selemone}}(k)
    \Big].
\end{split}\end{equation}

Finally, let $\paliifconc{i}$ ($\paliifres{i}{3}$) be the probability that Alice sent the signal $\cowsig{i}$, given that Eve's measurement was conclusive (inconclusive). These values can be written as
\begin{equation}\begin{split}
    \paliifconc{i} &=
    \frac{ \palice{i} \left( 1 - \presifali{3}{i} \right) }
    {\pconc},
    \\
    \paliifres{i}{3} &=
    \frac{ \palice{i} \presifali{3}{i} }
    {1 - \pconc},
\end{split}\end{equation}
for $i \in \{0,1,2\}$. Then, \cref{sec:SM_metrics} shows that
\begin{equation}\begin{split}
    &\faveofvisclks{1}{\selemtwo\selemone}(k) =
    \\
    &\frac{ \palice{\selemone} \palice{\selemtwo} }
    { \pconc^2 } \bigg\{
    (k-1) \Big[
    \left( 1 - \presifali{3}{\selemone} \right) \left( 1 - \presifali{3}{\selemtwo} \right)
    \\
    &- \presifalivis{0}{1} \left( 1 - \darkcount{\detectormonit{1}} \right)
    \Big]
    \\
    &+
    (k-1) \constrecursion^{k-2} \left( 1 - \darkcount{\detectormonit{1}} \right) \presifalivis{2}{2}
    \\
    &- \frac{1-\constrecursion^{k-1}}{1-\constrecursion} \left( 1 - \darkcount{\detectormonit{1}} \right) \Big[
    \left( 1 - \presifali{3}{\selemone} \right) \presifali{2}{\selemtwo}
    \\
    &+
    \presifali{2}{\selemone} \left( 1 - \presifali{3}{\selemtwo} \right)
    \Big]
    \bigg\},\\\\
\end{split}\end{equation}
and
\begin{equation}\begin{split}
    \faveofvisclks{2}{\selemtwo\selemone}(k) \!=&
    \frac{\palice{\selemone}\palice{\selemtwo}}
    {\pconc^2} \bigg\{
    (k-1) \Big[
    \presifalivis{0}{0}
    \\
    +& \presifalivis{0}{2}
    \!\!+ \presifalivis{1}{1}
    \\
    +& \presifalivis{2}{1}
    \!\!+ \darkcount{\detectormonit{2}} \Big(
    \presifalivis{0}{1}
    \\
    +& \presifalivis{1}{0}
    \!\!+ \presifalivis{1}{2}
    \\
    +& \presifalivis{2}{0}
    \!\!+ \presifalivis{2}{2} \Big) 
    \Big]
    \\
    -& \frac{1-\constrecursion^{k-1}}{1-\constrecursion}
    \left( 1 - \darkcount{\detectormonit{2}} \right)
    \!\Big( \presifalivis{0}{2}
    \\
    +& \presifalivis{2}{1} \Big)
    \bigg\},
\end{split}\end{equation}
while the parameters $\plastev{\eventvisclk{X}{\selemtwo\selemone}}(k)$ and $\pedgeev{\eventvisclk{X}{\selemtwo\selemone}}(k)$, for $X\!\in\!\set{1,2}$, are given by
\begin{equation}\label{eq:plastedge_vis}\begin{split}
    &\plastev{\eventvisclk{X}{\selemtwo\selemone}} (k)
    \\
    &= \begin{cases}
        0
        &\textrm{if } k = 0,
        \\
        \paliifconc{\selemone} \paliifres{\selemtwo}{3} \darkcount{\detbmonitlabel[X]}
        &\textrm{if } 0 < k < \maxblocklen,
        \\
        \paliifconc{\selemone} \palice{\selemtwo} \darkcount{\detbmonitlabel[X]}
        &\textrm{if } k = \maxblocklen,
    \end{cases}
    \\ 
    &\pedgeev{\eventvisclk{X}{\selemtwo\selemone}} (k)
    \\
    &= \begin{cases}
        \paliifres{\selemone}{3} \palice{\selemtwo} \darkcount{\detbmonitlabel[X]} 
        &\textrm{if } 0 \leq k < \maxblocklen,
        \\
        \palice{\selemone} \palice{\selemtwo} \darkcount{\detbmonitlabel[X]}
        &\textrm{if } k = \maxblocklen.
    \end{cases}
\end{split}\end{equation}

Finally, we note that $\visave$, as defined in \cref{eq:visave}, can be directly obtained from the previous averages $\aveofvisclks{X}{\svector}$ as follows
\begin{equation}
    \visave =
    \frac{
    \sum_{\svector} \aveofvisclks{1}{\svector} - \sum_{\svector} \aveofvisclks{2}{\svector}
    }{
    \sum_{\svector} \aveofvisclks{1}{\svector} + \sum_{\svector} \aveofvisclks{2}{\svector}
    }
\end{equation}
where the sums run for $\svector \in \{2,01,02,21,22\}$.

\section{Simulation results}\label{sec:results}

In this section we now evaluate the feasibility of the zero-error attack analyzed above in realistic conditions.
We shall denote the attack that uses the optimal USD measurement discussed in \cref{sec:implementation} as USD1.
Here, we shall include as well the results derived in~\cref{sec:USDM2} regarding an alternative, suboptimal USD measurement for Eve, which is able to discriminate not only data signals but also decoy signals, although its success probability is lower than that of USD1.
We shall denote this second strategy as USD2.
We refer the readers to \cref{sec:USDM2} for more details.
As it is shown below, USD2 can outperform USD1 in the presence of imperfections.

We consider that Eve's attack is successful if she can keep the resulting QBER and visibilities within certain acceptance intervals.
For illustration purposes, we shall use $\maxqber=0.05$ and $\minvis=0.95$ as the threshold values defining these acceptance intervals, \textit{i.e.}, the \qber and the visibilities $\visibility{\svector}$ must satisfy $\qber\leq\maxqber$ and $\visibility{\svector}\geq\minvis$ for the attack to be successful.
We remark, however, that these values are just an example chosen for continuity with previous studies~\cite{Gonzalez_Bounds:2020}, which in turn selected them to reflect the metrics attainable by state-of-the-art experiments~\cite{Stucki_HighRate:2009,Korzh_Provably:2015}.
In any case, our analysis can be applied to any other threshold values used to calculate the secret-key rate by considering a specific security proof.
Importantly, some commercial systems only check the average visibility $\visave$ out of all the visibilities~\cite{Korzh_Provably:2015,ID_Quantique}.
Below we show that this provides Eve a crucial advantage for her attack.

\begin{table}
    \begin{ruledtabular}
    \begin{tabular}{ccccccccc}
    $f$ & $|\alpha|^2$ & $\maxblocklen$ & $\phi$ & $T$ & $\amperror$ & $\effeve$ & $\darkcount{}$
    \\ \hline
    0.155 & 0.1 & 10 & $1^\circ$ & 0.99 & 0.05 & 0.6 & $10^{-7}$
    \end{tabular}
    \end{ruledtabular}
    \caption{Protocol and experimental parameters used in the simulations.
    We consider the same dark-count probability $\darkcount{}$ for all detectors, \textit{i.e.}, $\darkcount{\detectoreve}= \darkcount{\detectordata}= \darkcount{\detectormonit{X}}=\darkcount{}$.}
    \label{tab:params}
\end{table}

\begin{figure}
  \includegraphics[width=1\columnwidth]{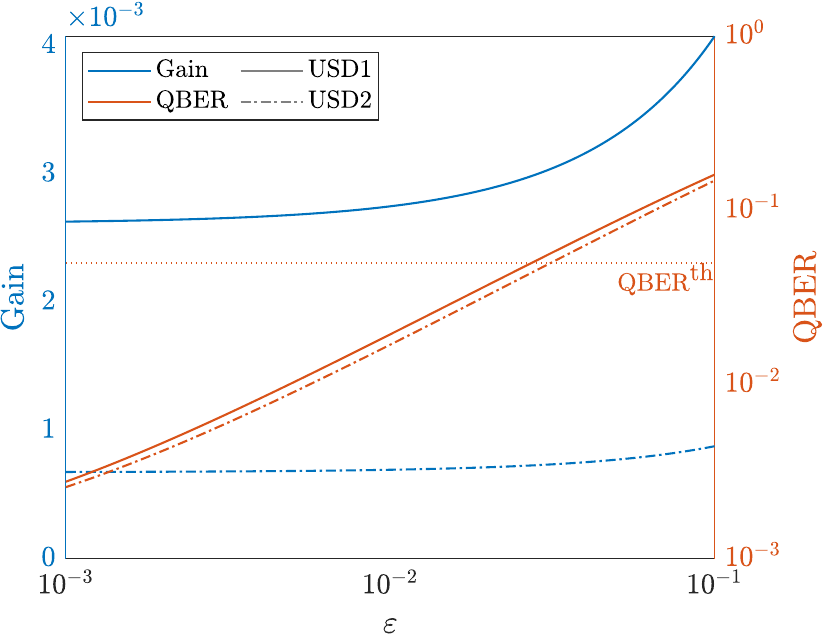}
  \caption{Resulting gain $G$ and QBER for a COW QKD system in the presence of Eve's attack as a function of the error parameter $\varepsilon$.
  This parameter models Eve's imperfect mode overlap in her measurement implementation. Solid lines correspond to USD1, while dashed lines correspond to USD2. For the simulations we considered the parameters given in~\cref{tab:params}.}
  \label{fig:metrics_gain_qber}
\end{figure}

\cref{fig:metrics_gain_qber} illustrates the resulting gain $\gain$ and $\qber$ as a function of an error parameter $\varepsilon$, which is directly related to the quality of the mode overlap at Eve's beamsplitters.
Specifically, we set $t_1=t_2=1-\varepsilon$ for USD1, and $t_1=t_2=t_3=t_4=1-\varepsilon$ for USD2 (see~\cref{sec:USDM2}).
Here we focus on the mode overlap because our results suggest that this is the main limiting experimental factor in Eve's attack, while its effectiveness varies only slightly when the experimental parameters that model other imperfections are changed over realistic ranges of values (see~\cref{sec:other_imperfections} for further details).
In particular, for the simulations in~\cref{fig:metrics_gain_qber}, we assign to other imperfections the values given in \cref{tab:params}.
For simplicity, we consider the same dark-count probability $\darkcount{}$ for all detectors, \textit{i.e.}, $\darkcount{\detectoreve}= \darkcount{\detectordata}= \darkcount{\detectormonit{X}}=\darkcount{}$, with $X\!\in\!\set{1,2}$.
Besides, we set the parameters of the COW protocol, namely $\alpha$ and $f$, to typical values, also shown in \cref{tab:params}, and we fix $\maxblocklen=10$, as we observe essentially no improvement of the metrics when increasing this value, given that the intensity of Alice's pulses is kept within practical margins (further details are discussed in \cref{sec:other_imperfections}).

Remarkably, \cref{fig:metrics_gain_qber} shows that the gain $\gain$ grows with $\varepsilon$.
To understand this, we note that, for low values of $|\alpha|^2$ and small imperfections, both USD measurements have a small probability of being conclusive. However, as imperfections escalate, the probability of a conclusive measurement increases slightly due to a slight rise in the click probability of Eve's detectors.
As expected, USD1 allows for a higher gain than USD2, as the former has been designed to maximize the probability of identifying data signals.
As for the QBER, the figure shows that Eve can keep it below the chosen threshold value of 0.05 for reasonably high values of $\varepsilon$.

\begin{figure}
    \vspace{2mm}
    \includegraphics[width=0.97\columnwidth]{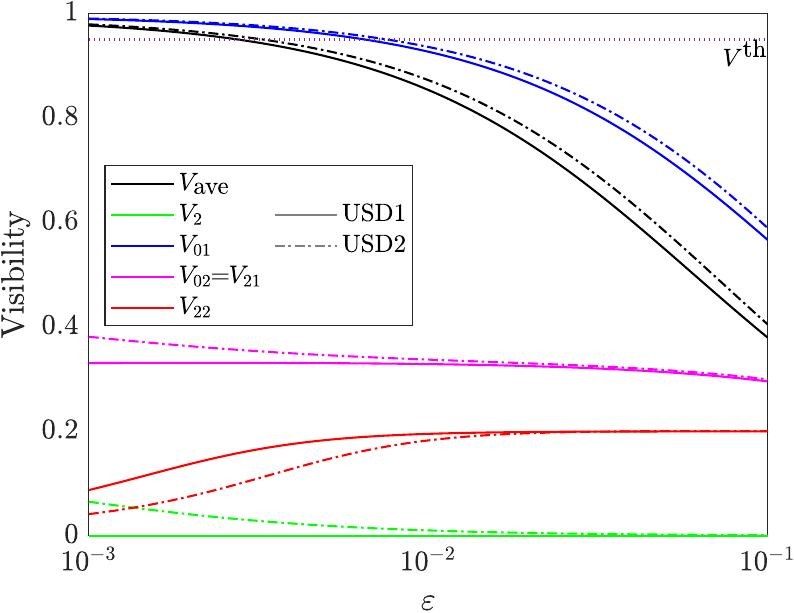}
    \caption{Resulting visibilities for a COW QKD system in the presence of Eve's attack as a function of the error parameter $\varepsilon$. This parameter models Eve's imperfect mode overlap in her measurement implementation. Solid lines correspond to USD1, while dashed lines correspond to USD2. For the simulations we consider the parameters given in~\cref{tab:params}.}
    \label{fig:metrics_visibs}
\end{figure}

The visibilities are illustrated in \cref{fig:metrics_visibs}. We observe that those corresponding to sequences containing a decoy signal are well below the threshold $\minvis$, being this particularly evident for $\visibility{2}$, which is in fact always zero for USD1.
This is expected, as USD1 is unable to identify $\cowsig{2}$.
Indeed, when receiving $\cowsig{2}$, Eve will mostly resend vacuum signals to Bob.
Certainly, due to measurement errors, she could occasionally misidentify Alice's signal and resend Bob $\cowsig{0}$ or $\cowsig{1}$.
However, these signals can trigger both detectors in Bob's monitoring line with equal probability, and consequently do not increase the visibility.
Needless to say, dark counts cannot increase the expected visibility either, as they occur with equal probability in both detectors.
Interestingly, even for USD2, the visibility $\visibility{2}$ is notably low.
The reason for this behavior is twofold: first, the probability that Eve identifies $\cowsig{2}$ is relatively low.
Second, the probability of obtaining a conclusive measurement outcome is also low for the considered value of $|\alpha|^2$, so most blocks processed by Eve are expected to be short.
Therefore, it is likely that those few $\cowsig{2}$ that are correctly identified are located at the edge of a block, where they are consequently erased due to Eve's block-processing strategy.

Similar arguments apply to $\visibility{22}$, $\visibility{21}$, and $\visibility{02}$ (which is equal to $\visibility{21}$ due to the symmetry of the setups).
In these cases, however, the visibilities are non-zero even for USD1.
This is because their corresponding sequences are occasionally translated by Eve to the sequence \sequence{01}, which positively contributes to the visibility (see~\cref{sec:explanation_visib} for further details).
Remarkably, we note that USD2 performs better than USD1 for most visibilities in this scenario.
This is because the probability of erroneously identifying $\cowsig{0}$ as $\cowsig{1}$ (or viceversa), given that Eve's measurement is conclusive, is smaller in USD2.
For the same reason, $\visibility{01}$ is also larger for USD2.
This is relevant because Alice typically transmits the sequence \sequence{01} much more frequently than the other sequences, as $f$ is usually set to a small value to increase the number of data rounds.
As a consequence, $\visibility{01}$, the highest visibility, is also the largest contributor to the average visibility $\visave$, which thus remains relatively high for reasonably small values of $\varepsilon$.

To compare the performance of Eve's attack with respect to prior studies~\cite{Gonzalez_Bounds:2020,Trenyi_Attack:2021}, we compute here a simple upper bound on the secret-key rate.
With this in mind, let us conveniently refer to Eve's attack as \textit{undetectable} if the following two conditions are met.
Firstly, both the QBER and the average visibility $\visave$ observed for the attacked system must fall within their corresponding acceptance regions, \textit{i.e.}, $\qber < \maxqber$ and $\visave > \minvis$.
Secondly, the gain $\gain$ of the attacked system must equal or exceed that expected from a legitimate system, which we will denote as $\gain^\legitlabel$ to indicate that no attack is launched.
In particular, in the simulations we consider a typical lossy channel model for which~\cite{Trenyi_Attack:2021}
\begin{equation}\label{eq:noattack_gain}\begin{split}
    \gain^\legitlabel =\:&
    1 - \Big[
    (1-f) e^{-\effbob\transmittance t_B |\alpha|^2}
    \\
    +\:& f e^{-2\effbob\transmittance t_B |\alpha|^2}
    \Big] \left( 1 - \darkcount{\detectordata} \right)^2,
\end{split}\end{equation}
where $\effbob$ is the efficiency of Bob's detectors, $t_B$ is transmittance of the beamsplitter used by Bob to separate between data and monitoring line (see \cref{fig:cow_scheme}), $\transmittance = 10^{-\attconstant d / 10}$ is the channel transmittance, $d$ is the channel distance (in km), and $\attconstant$ is the attenuation coefficient (in dB/km).
Importantly, if for a given $\transmittance$ Eve's attack meets the previous two conditions ---\textit{i.e.}, if it is undetectable--- it immediately follows that no secret key can be distilled by Alice and Bob based on the observed metrics.
Indeed, the secret-key rate $K$ can be simply upper bounded as~\cite{Gonzalez_Bounds:2020,Trenyi_Attack:2021}
\begin{equation}\label{eq:keyrate}
    K < (1-f) \transmittance \effbob \mu_\textrm{max} \equiv K_\textrm{max},
\end{equation}
where $\mu_\textrm{max}$ is the maximum value of $\mu$ for which Eve cannot perform an undetectable attack. We remark that, in general, $\mu_\textrm{max}$ depends on $\transmittance$. For example, for long distances $\gain^\legitlabel$ is expected to be low, and so it is easier for Eve to guarantee the condition $G\geq\gain^\legitlabel$. However, one could compensate the low $\gain^\legitlabel$ by reducing $\mu$ to make Alice's signals harder to identify, thereby reducing $G$.
That is, $\mu_\textrm{max}$ typically decreases with the channel distance.
In particular, previous works~\cite{Gonzalez_Bounds:2020,Trenyi_Attack:2021} showed that $\mu_\textrm{max}$ scales linearly with $\transmittance$ when no technological constrains are placed on Eve. Since $\transmittance$ already appears in~\cref{eq:keyrate}, the resulting scaling of $K_\textrm{max}$ is quadratic with $\transmittance$.

\begin{figure}
    \centering
    \includegraphics[width=1\columnwidth]{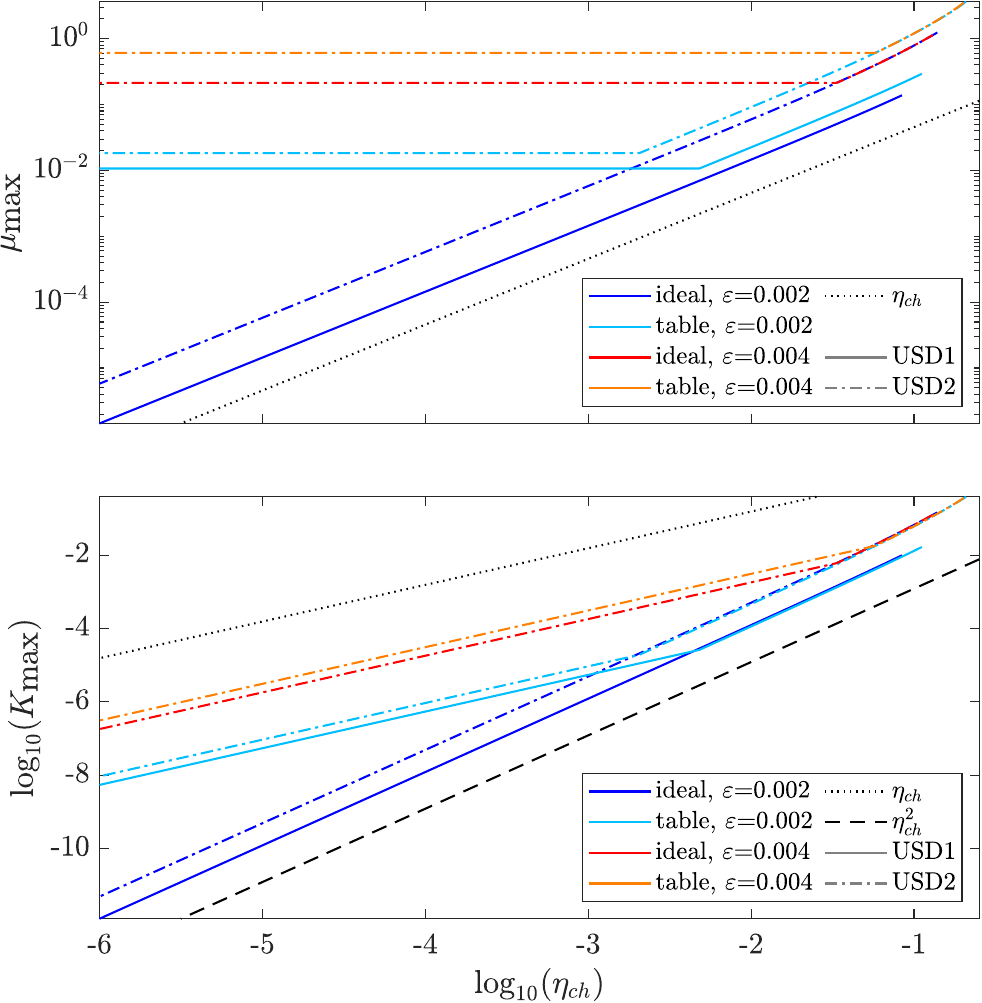}
    \caption{Maximum intensity that Alice can use while still being able to detect Eve's attack, $\mu_\textrm{max}$, and upper bound of the key rate, $K_\textrm{max}$, for different channel transmittances.
    The lines labeled as \textit{ideal} in the legend were computed by considering no imperfections in Eve's measurement besides the mode mismatch characterized by $\varepsilon$, while those labeled \textit{table} used the parameters in \cref{tab:params}.
    The dotted (dashed) black line is added as a reference to show linear (quadratic) scaling with $\transmittance$.\label{fig:keyrate_bounds}
    }
\end{figure}

To investigate if a technologically constrained eavesdropper can impose the same restrictive scaling, we plot $\mu_\textrm{max}$ and $K_\textrm{max}$ against $\transmittance$ in \cref{fig:keyrate_bounds}.
We consider two scenarios, which we evaluate for two different values of $\varepsilon$ each, and for both USD1 and USD2.
In the first scenario, we assume that all of the devices used by Eve and the legitimate users are ideal, and the only imperfection is the mode mismatch characterized by $\varepsilon$.
The second scenario considers the practical parameters introduced in \cref{tab:params}, and sets $\effbob=\effeve=0.6$ (notice that this is a rather conservative assumption, since Eve's technological capabilities are expected to surpass those of the legitimate users).
We set $f=0.155$ and $t_B=0.9$ in both instances.
In terms of imperfect mode overlap, we run the simulations for $\varepsilon=0.002$, which is sufficiently low to guarantee that $\visave > \minvis$ for both USD1 and USD2 and typical values of $\alpha$ (see \cref{fig:metrics_visibs}), and for $\varepsilon=0.004$, which results in a $\visave$ slightly below $\minvis$.
The results are plotted for a range of $\transmittance$ that corresponds to channel lengths between 30 and 300 km when considering a typical fiber-loss coefficient $\attconstant=0.2$ dB/km.

In line with~\cite{Gonzalez_Bounds:2020,Trenyi_Attack:2021}, a quadratic scaling across the entire range of $\transmittance$ is observed for both the USD1 and USD2 when considering the ideal parameters and $\varepsilon=0.002$.
However, this is no longer the case when considering the parameters from \cref{tab:params}, for which the quadratic scaling is only observed for up to $\transmittance \approx 10^{-2.5}$.
When a smaller $\transmittance$ ---\textit{i.e.}, longer channel--- is considered, $\mu_\textrm{max}$ becomes so small that Eve's attack can no longer fulfill the condition $\visave > \minvis$.
Therefore, as there is no need to reduce $\gain$ to prevent Eve's attack from being undetectable, $\mu_\textrm{max}$ remains constant from there on, and the scaling of $K_\textrm{max}$ turns from quadratic to linear over $\transmittance$.
Interestingly, even though the upper bound imposed by the USD1 is stricter than that imposed for the USD2, the latter remains in the quadratic regime for longer.
This is because the performance of the USD2 is worse in terms of gain, but better in terms of visibility (see~\cref{fig:metrics_gain_qber,fig:metrics_visibs}).

Moving to $\varepsilon=0.004$, we see that the USD2 barely allows to maintain the upper bound in the quadratic regime, and indeed the scaling is linear for $\transmittance \gtrapprox 10^{-1.5}$.
Naturally, larger errors in the implementation imply that Eve requires a higher value of $\mu$ to successfully attack the system.
Therefore, by keeping $\mu$ below this level, it is guaranteed that Eve's attack is detectable.
Moreover, the implementation using USD1 is completely unable to impose a bound on the secret-key rate at any distance.
This is because the metrics under attack improve with $\mu$, but only up to a certain optimal point. This point may appear at a different value of $\mu$ for each metric, and is a consequence of several factors (\textit{e.g.}, very high values of $\mu$ may result in lots of erroneous clicks at Eve's detectors, preventing her attack from remaining undetectable).
This means that if for one of these optimal intensities the metrics do not all fall within their acceptance regions, then the attack will be unsuccessful at any other intensity, and so $\mu_\textrm{max}\to\infty$ and $K_\textrm{max}\to\infty$.

\begin{figure}
  \includegraphics[width=1\columnwidth]{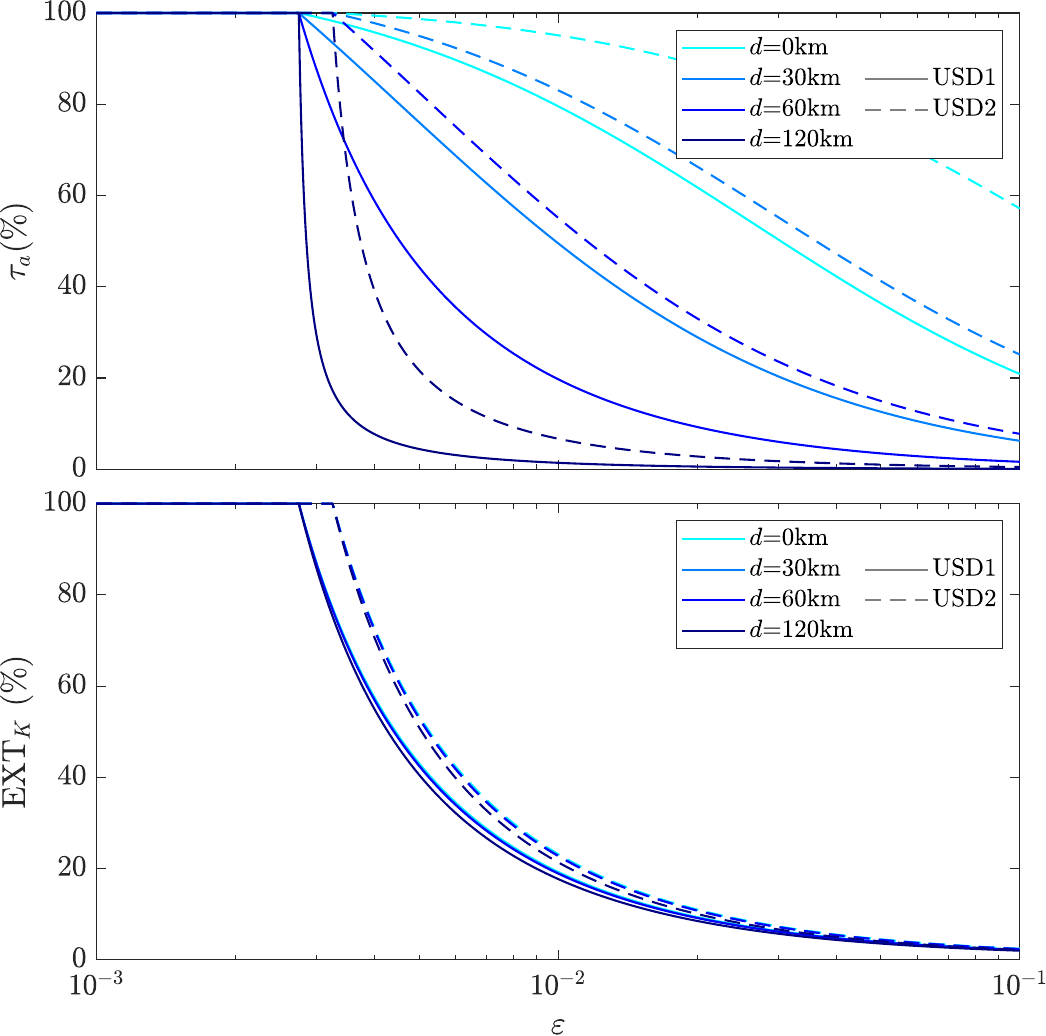}
  \caption{
  Fraction $\attackratio$ (in percentage) of the rounds that can be attacked by Eve while maintaining $\qber\leq 0.05$ and $\visave\geq0.95$, and percentage of sifted key $\text{EXT}_K$ that Eve can extract in such case, as a function of the error parameter $\varepsilon$ and the transmission distance $d$ between Alice and Bob.
  The parameter $\varepsilon$ models Eve's imperfect mode overlap in her measurement implementation.
  For the simulations we set the attenuation coefficient of the channel to $\attconstant=0.2$dB/km and consider the parameters given in~\cref{tab:params}, as well as the efficiency of Bob's detectors to be equal to that of Eve's.
  Note that the plots for the cases $d$=30km and $d$=60km in the right graph are almost overlapped for both USD1 and USD2, with the plot for $d$=0km just above them with some amount of overlap too.
  }
  \label{fig:attackable_key}
\end{figure}

It is clear that, according to our simple model, achieving a precise mode overlap within the beamsplitters is crucial to the success of Eve's attack.
Nonetheless, an eavesdropper encountering challenges in this regard might still manage to extract some amount of information by launching a partial attack, in which she acts only on a reduced subset of the transmitted signals. 
\Cref{fig:attackable_key} illustrates, for different values of the channel distance $d$, the fraction $\attackratio$ of the rounds that can be attacked by Eve while still maintaining her success with respect to the QBER and $\visave$, as well as the maximum percentage of sifted key that she can extract in this scenario (see \cref{sec:sparse_attack} for details about how these quantities are computed).
In particular, this means that both metrics are maintained within their corresponding acceptance ranges.
Here, we assume a typical fiber-based channel with attenuation coefficient $\attconstant=0.2$ dB/km.
\cref{fig:attackable_key} shows that a shorter channel length results in a higher amount of rounds that can be attacked by Eve.
This is to be expected, since the metrics are calculated from all the rounds, those intercepted by Eve and those that are not. 
Consequently, the better measurement statistics observed by Bob at short distances from the unattacked rounds favourably impact the overall value of the metrics.
Moreover, \cref{fig:attackable_key} also shows that the number of rounds that Eve may attack falls very sharply for longer channel lengths once the metrics of her attack are outside of the acceptability ranges.
Again, this is due to the degradation of the legitimate users' metrics, as it is clear that in the extreme case where the $\qber$ and visibility obtained during the unattacked rounds are equal to the threshold values $\maxqber$ and $\minvis$, no additional error is permissible, and so Eve cannot attack any rounds without becoming detectable unless her metrics are an improvement over those of Alice and Bob.
Nevertheless, the amount of sifted-key bits distilled during the unattacked rounds also increases at short distances, meaning that the ratio of sifted key known to Eve decreases.
For the scenario described by the parameters in \cref{tab:params}, these two opposite effects result in very little variation of $\text{EXT}_K$ over the distance $d$, as can be seen at the right in \cref{fig:attackable_key}.
Crucially, the figure shows that Eve's attack has the potential to compromise the entire secret key at low values of $\varepsilon$.
What is more, even for higher values of $\varepsilon$, Eve can still successfully attack an important fraction of the rounds and obtain part of the sifted key.

\section{Conclusions}\label{sec:conclusions}
Security proofs of quantum key distribution (QKD) typically consider that the eavesdropper’s capabilities are only limited by the laws of quantum mechanics.
Here, we have evaluated a less conservative scenario in which Eve is actually restricted by current technology.
In particular, we have studied the feasibility of zero-error attacks against coherent-one-way (COW) QKD in this framework.
To do so, we have introduced two practical receivers to perform an unambiguous state discrimination (USD) measurement of Alice’s emitted signals, which is the essential step of this type of attack.

Both proposed USD receivers are rather simple, and employ only linear passive optics, phase-space displacement operations and threshold single-photon detectors.
We have derived analytical expressions for the expected values of the main metrics (\textit{i.e.}, the gain, the quantum bit error rate and the visibilities) of a COW QKD protocol assuming realistic experimental conditions, \textit{i.e.}, as a function of the most relevant device imperfections of Eve’s equipment.
In doing so, we have found that the most critical experimental parameter seems to be the quality of interference between Alice’s weak coherent pulses and Eve’s strong light during her displacement operation.
Overall, our results indicate that zero-error attacks could break the security of COW QKD with present-day technology, particularly if Alice and Bob only consider the observed average visibility in their monitoring line, as it is done \textit{e.g.} in commercial setups and long-distance implementations of this scheme.

\section{Acknowledgements}\label{sec:acknowledgements}
This work was supported by Cisco Systems Inc., the Galician Regional Government (consolidation of Research Units: AtlantTIC), the Spanish Ministry of Economy and Competitiveness (MINECO), the Fondo Europeo de Desarrollo Regional (FEDER) through the grant No. PID2020-118178RB-C21, MICIN with funding from the European Union NextGenerationEU (PRTR-C17.I1) and the Galician Regional Government with own funding through the “Planes Complementarios de I+D+I con las Comunidades Autónomas” in Quantum Communication, the European Union’s Horizon Europe Framework Programme under the Marie Sklodowska-Curie Grant No. 101072637 (Project QSI), as listed by the UKRI though the Engineering and Physical Sciences Research Council (EPSRC) with grant No. EP/X028313/1, and the project “Quantum Security Networks Partnership” (QSNP, grant agreement No. 101114043).
P.v.L. acknowledges funding from the BMBF in Germany (QR.X and QuaPhySI) and from the EU/BMBF via QuantEra (ShoQC).

\section{Data availability statement}
All data that support the findings of this study are included within the article (and any supplementary files).

\appendix
\section{Model for the imperfect mode overlap}\label{sec:mode_mismatch}

\begin{figure}
    \centering
    \includegraphics[width=0.8\linewidth]{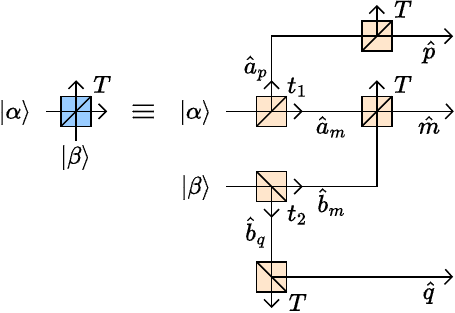}
    \caption{Simple model to characterize the imperfect mode overlap at one output of a beamsplitter with two coherent pulses $\ket{\alpha}$ and $\ket{\beta}$ as inputs.
    The original (blue) beamsplitter of transmittance $T$ at the left, where pulses $\ket{\alpha}$ and $\ket{\beta}$ do not perfectly interfere, can be modelled as the network of perfectly mode-matched (orange) beamsplitters shown at the right. In this network, the input state $\ket{\alpha}$ ($\ket{\beta}$) is first split into two optical modes, $\hat{a}_m$ and $\hat{a}_p$ ($\hat{b}_m$ and $\hat{b}_q$), according to the parameter $t_1$ ($t_2$) that describes the corresponding beamsplitter transmittance. Subsequently, the optical modes $\hat{a}_m$ and $\hat{b}_m$ interfere in a beamsplitter with transmittance $T$, while each of the modes $\hat{a}_p$ and $\hat{b}_q$ interferes with vacuum in a beamsplitter with transmittance also $T$, resulting in the final output modes of the network $\hat{p}$, $\hat{m}$ and $\hat{q}$.}
    \label{fig:laiho_model}
\end{figure}

Let us consider a real beamsplitter with transmittance $T$ in which the two input optical pulses are not perfectly mode-matched, and therefore do not interfere perfectly (see \cref{fig:laiho_model}).
This imperfect mode overlap can be simplistically modelled by splitting each of the two inputs into two modes~\cite{Laiho_Probing:2009}.
One of these modes from each pulse interferes as desired, while the two remaining modes do not interfere and only have their amplitudes diminished by the beamsplitter.
The fraction of light from the first (second) input that is perfectly matched and therefore interferes in the beamsplitter is determined by the parameter $t_1$ ($t_2$), and the total degree of overlap in the beamsplitter is  defined as $\mathcal{M} := t_1t_2$~\cite{Laiho_Probing:2009}.

In order to model the splitting of the first (second) input into two modes, namely one that is properly matched, say $\hat{a}_m$ ($\hat{b}_m$), and one that does not interfere, say $\hat{a}_p$ ($\hat{b}_q$), an ideal beamsplitter of transmittance $t_1$ ($t_2$) can be used, as illustrated in \cref{fig:laiho_model}.
After that, the modes that are properly matched (\textit{i.e.}, $\hat{a}_m$ and $\hat{b}_m$) interfere in an ideal beamsplitter with the original transmittance $T$, resulting in a new mode, say $\hat{m}$, at one output of the beamsplitter.
The other modes (i.e. $\hat{a}_p$ and $\hat{b}_q$) interfere with vacuum at ideal beamsplitters with transmittance $T$, resulting in new modes, say $\hat{p}$ and $\hat{q}$, at one output of each beamsplitter.

\section{Derivation of the metrics}\label{sec:SM_metrics}

\Cref{sec:metrics} of the main text provides all the necessary expressions to compute the expected values of the metrics for a system in the presence of Eve, given the configuration of the involved parties.
It also includes a step-by-step derivation of the gain, $\gain$.
This Appendix provides the most relevant intermediate results for the derivation of both the QBER and visibilities, alongside with guidance on the interpretation of certain values.

\subsection{QBER}
Here we derive analytical expressions for the quantities $\faveofdataclks(k)$ and $\faveoferrors(k)$ presented in \cref{eq:n_dataclk,eq:n_error}. The derivation of \cref{eq:plast_dataclk,eq:plast_error} is straightforward.

First, we note that both $\faveofdataclks(k)$ and $\faveoferrors(k)$ admit a decomposition similar to that used for $\faveofclks(k)$ in~\cref{eq:n_click_k}.
That is, we can express
\begin{equation}
    \begin{split}
        \faveofdataclks(k) &= \sum_{i=0}^2 \presifconc{i} \faveofdataclks(k|i),\\
        \faveoferrors(k) &= \sum_{i=0}^2 \presifconc{i} \faveoferrors(k|i),
    \end{split}
\end{equation}
where $\faveofdataclks(k|i)$ ($\faveoferrors(k|i)$) denotes the average number of key bits (erroneous key bits) distilled from a conclusive-block of length $k$ given that the first signal in the block is identified by Eve as $\cowsig{i}$.

Let $\alice_\textrm{key}$ represent the event in which Alice emits a data signal, \textit{i.e.}, $\alice_\textrm{key} = \alice_0\!\cup\!\alice_1$.
Then, it follows that $\palidata = \palice{0} + \palice{1}$, $\pdataifres{j} = \paliifres{0}{j} + \paliifres{1}{j}$, and $\pdataifconc = \paliifconc{0} + \paliifconc{1}$.
Additionally, let $\perrifbob{i}$ be the probability that Bob distills an incorrect key bit (\textit{i.e.}, a key bit different from the one sent by Alice), given that he receives $\cowsig{i}$ from Eve.
Note that this implies that the result of Eve's measurement for that signal was $\eve_i$ and this signal was not discarded by her processing.
Moreover, we define $\perrifbobnvac = \sum_{i=0}^2 \perrifbob{i}$ as the probability that Bob distills an erroneous bit given that he received a signal different from $\vacsig$.
Then, we have that
\begin{equation}\begin{split}
    &\perrifbob{0} =
    \paliifres{0}{0} \frac{\darkcount{\detectordata}}{2}
    + \paliifres{1}{0} \left( 1 - \frac{\darkcount{\detectordata}}{2} \right),
    \\
    &\perrifbob{1} =
    \paliifres{0}{1} \left( 1 - \frac{\darkcount{\detectordata}}{2} \right)
    + \paliifres{1}{1} \frac{\darkcount{\detectordata}}{2},
    \\
    &\perrifbob{2} =
    \frac{ \paliifres{0}{2} + \paliifres{1}{2} }{2}.
\end{split}\end{equation}
With this, we can express both $\faveofdataclks(k|i)$ and $\faveoferrors(k|i)$ in a recursive form. Specifically, we have that
\begin{equation}\begin{split}
    \faveofdataclks(k|0) =\:&
    \presifconc{0} \big[
    \pdataifres{0} \left( 1 + \pclickifvac \right)
    \\
    +\:& \pdataifconc (k-2)
    \big]
    \\
    +\:&
    \presifconc{1} \big[
    \left( \pdataifres{0} + \pdataifres{1} \right) \pclickifvac
    \\
    +\:& \pdataifconc (k-2)
    \big]
    \\
    +\:&
    \presifconc{2} \big[ \faveofdataclks(k-1|0)
    \\
    +\:& \pdataifres{2} \pclickifvac \big],
    \\ 
    \faveofdataclks(k|1) =\:&
    \presifconc{0} \big[
    \pdataifres{0} + \pdataifres{1}
    \\
    +\:& \pdataifconc (k-2)
    \big]
    \\
    +\:&
    \presifconc{1} \big[
    \pdataifres{1} \left( 1 + \pclickifvac  \right)
    \\
    +\:& \pdataifconc (k-2)
    \big]
    \\
    +\:&
    \presifconc{2} \big[ \faveofdataclks(k-1|1) 
    \\
    +\:& \pdataifres{2} \pclickifvac \big],
    \\ 
    \faveofdataclks(k|2) =\:&
    \faveofdataclks(k-1) + \pdataifres{2} \pclickifvac,
\end{split}\end{equation}
and
\begin{equation}\begin{split}
    \faveoferrors(k|0) =\:&
    \presifconc{0} \big[
    \pdataifres{0} \perrorifvac + \perrifbob{0}
    \\
    +\:& \perrifbobnvac (k-2)
    \big]
    \\
    +\:&
    \presifconc{1} \big[
    \left( \pdataifres{0} + \pdataifres{1} \right) \perrorifvac
    \\
    +\:& \perrifbobnvac (k-2)
    \big]
    \\
    +\:&
    \presifconc{2} \big[
    \faveoferrors(k-1|0)
    \\
    +\:& \pdataifres{2} \perrorifvac
    \big],
    \\ 
    \faveoferrors(k|1) =\:&
    \presifconc{0} \big[
    \perrifbob{0} + \perrifbob{1}
    \\
    +\:& \perrifbobnvac (k-2)
    \big]
    \\
    +\:&
    \presifconc{1} \big[
    \perrifbob{1} + \pdataifres{1} \perrorifvac
    \\
    +\:& \perrifbobnvac (k-2)
    \big]
    \\
    +\:&
    \presifconc{2} \big[
    \faveoferrors(k-1|1)
    \\
    +\:& \pdataifres{2} \perrorifvac
    \big],
    \\ 
    \faveoferrors(k|2) =\:&
    \faveoferrors(k-1) + \pdataifres{2} \perrorifvac.
\end{split}\end{equation}
 The starting points for the recursions above are
\begin{equation}
    \begin{split}
        \faveofdataclks(1|i) =\:&
        \pdataifres{i} \pclickifvac,
        \\
        \faveoferrors(1|i) =\:&
        \pdataifres{i} \perrorifvac,
    \end{split}
    \qquad
    \begin{split}
        \faveofdataclks(0) =\:& 0,
        \\
        \faveoferrors(0) =\:& 0,
    \end{split}
\end{equation}
for $i \in \{0,1\}$.
By solving the recursions, one finally obtains~\cref{eq:n_dataclk,eq:n_error}.

\subsection{Visibilities}
We focus first on the visibility $\visibility{2}$.
According to \cref{eq:visibilities} and \cref{eq:aveofvis2}, this visibility can be computed from the values of $\faveofvisclks{X}{2}(k)$ and $\plastev{\eventvisclk{X}{2}}(k)$, with $X \in \{1,2\}$.
It is straightforward to obtain the value of $\plastev{\eventvisclk{X}{2}}(k)$, shown in \cref{eq:plast_vis2}, so we will focus here on $\faveofvisclks{X}{2}(k)$.

Let $\presifalitoconc{i}{j}$ denote the conditional probability that the outcome of Eve's USD measurement is $\eve_i$, given that Alice prepared the signal $\cowsig{j}$ and Eve's measurement was conclusive. That is,
\begin{equation}
    \presifalitoconc{i}{j} =
    \frac{ \presifali{i}{j} }{ \presifali{0}{j} + \presifali{1}{j} + \presifali{2}{j} },
\end{equation}
for $i,j \in \{0,1,2\}$.
Now, we express $\faveofvisclks{X}{2}(k)$ in the form 
\begin{equation}
    \faveofvisclks{X}{2}(k) = \sum_{i=0}^2 \presifconc{i} \faveofvisclks{X}{2}(k|i),
\end{equation}
where $\faveofvisclks{X}{2}(k|i)$ is the average number of signals sent by Alice as $\cowsig{2}$ that prompt a click in $\detbmonit[X]$ within a conclusive-block of length $k$, given that the first signal of the block is $\cowsig{i}$. 
We can write recursive expressions to describe these quantities, such as
\begin{equation}\begin{split}
    \faveofvisclks{1}{2}(k|0) =\:&
    \presifconc{0} \big[
    \paliifres{2}{0} \left( 1 + \darkcount{\detectormonit{1}} \right)
    \\
    +\:& \paliifconc{2} (k-2)
    \big]
    \\
    +\:&
    \presifconc{1} \big[
    \left( \paliifres{2}{0} + \paliifres{2}{1} \right) \darkcount{\detectormonit{1}}
    \\
    +\:& \paliifconc{2} (k-2)
    \big]
    \\
    +\:&
    \presifconc{2} \big[
    \faveofvisclks{1}{2}(k-1|0)
    + \paliifres{2}{2} \darkcount{\detectormonit{1}}
    \big],
    \\ 
    \faveofvisclks{1}{2}(k|1) =\:&
    \presifconc{0} \big[
    \paliifres{2}{0} + \paliifres{2}{1}
    \\
    +\:& \paliifconc{2} (k-2)
    \big]
    \\
    +\:&
    \presifconc{1} \big[
    \paliifres{2}{1} \left( 1 + \darkcount{\detectormonit{1}} \right)
    \\
    +\:& \paliifconc{2} (k-2)
    \big]
    \\
    +\:&
    \presifconc{2} \big[
    \faveofvisclks{1}{2}(k-1|1)
    + \paliifres{2}{2} \darkcount{\detectormonit{1}}
    \big],
    \\ 
    \faveofvisclks{1}{2}(k|2) =\:&
    \faveofvisclks{1}{2}(k-1) + \paliifres{2}{2} \darkcount{\detectormonit{1}},
\end{split}\end{equation}
and
\begin{equation}\begin{split}
    \faveofvisclks{2}{2}(k|0) =\:&
    \presifconc{0} \big[
    \paliifres{2}{0} \left( 1 + \darkcount{\detectormonit{2}} \right)
    \\
    +\:& \paliifconc{2} \left( 1 - \presifalitoconc{2}{2} \right) (k-2)
    \big]
    \\
    +\:&
    \presifconc{1} \big[
    \left( \paliifres{2}{0} + \paliifres{2}{1} \right) \darkcount{\detectormonit{2}}
    \\
    +\:& \paliifconc{2} \left( 1 - \presifalitoconc{2}{2} \right) (k-2)
    \big]
    \\
    +\:&
    \presifconc{2} \big[
    \faveofvisclks{2}{2}(k-1|0)
    + \paliifres{2}{2} \darkcount{\detectormonit{2}}
    \big],
    \\ 
    \faveofvisclks{2}{2}(k|1) =\:&
    \presifconc{0} \big[
    \paliifres{2}{0} + \paliifres{2}{1}
    \\
    +\:& \paliifconc{2} \left( 1 - \presifalitoconc{2}{2} \right) (k-2)
    \big]
    \\
    +\:&
    \presifconc{1} \big[
    \paliifres{2}{1} \left( 1 + \darkcount{\detectormonit{2}} \right)
    \\
    +\:& \paliifconc{2} \left( 1 - \presifalitoconc{2}{2} \right) (k-2)
    \big]
    \\
    +\:&
    \presifconc{2} \big[
    \faveofvisclks{2}{2}(k-1|1)
    + \paliifres{2}{2} \darkcount{\detectormonit{2}}
    \big],
    \\ 
    \faveofvisclks{2}{2}(k|2) =\:&
    \faveofvisclks{2}{2}(k-1) + \paliifres{2}{2} \darkcount{\detectormonit{2}},
\end{split}\end{equation}
being the starting points of the recursions
\begin{equation}\begin{split}
    \faveofvisclks{X}{2}(1|i) =&
    \paliifres{2}{i} \darkcount{\detectormonit{X}},
    \qquad
    \faveofvisclks{X}{2}(0) = 0,
\end{split}\end{equation}
for $X \in \{1,2\}$ and $i \in \{0,1\}$.
By solving the recursions, one obtains~\cref{eq:n_M12_M22}.

Now we focus on the visibilities $\visibility{\svector}$ of sequences $\svector = \selems$ that contain two signals.
From \cref{eq:avefromfave_vis}, we have that these visibilities can be computed from the quantities $\faveofvisclks{X}{\selems}(k)$, $\plastev{\eventvisclk{X}{\selems}}(k)$ and $\pedgeev{\eventvisclk{X}{\selems}}(k)$, introduced in \cref{sec:metrics_vis}.
The quantities $\plastev{\eventvisclk{X}{\selems}}(k)$ and $\pedgeev{\eventvisclk{X}{\selems}}(k)$ are given in \cref{eq:plastedge_vis}, and their derivation is straightforward.
Thus we focus on the derivation of $\faveofvisclks{X}{\selems}(k)$.

Following a similar approach as with the previous metrics, first we write
\begin{equation}
    \faveofvisclks{X}{\selems}(k) = \sum_{i=0}^2 \presifconc{2}\faveofvisclks{X}{\selems}(k|i),
\end{equation}
and then we focus on finding the quantities $\faveofvisclks{X}{\selems}(k|i)$.
Now, let $\pcohclick{X}$ denote the conditional probability that Alice originally prepares the sequence $\selemtwo\selemone$ and $\detbmonit[X]$ registers a click in the time slot associated with the interference between the last pulse of $\selemone$ and the first pulse of $\selemtwo$, given that Bob received a non-vacuum signal in both rounds.
This implies that Eve measured both signals conclusively and they were not discarded during her processing.
Then we have that
\begin{equation}\begin{split}
    \pcohclick{1} =\:&
    \paliifconc{\selemone} \paliifconc{\selemtwo} \Big[
    1
    \\
    -\:& \presifalitoconc{0}{\selemone} \presifalitoconc{1}{\selemtwo} \left( 1 - \darkcount{\detectormonit{1}} \right)
    \Big],
\end{split}\end{equation}
and
\begin{equation}\begin{split}
    \pcohclick{2} =\:&
    \paliifconc{\selemone} \paliifconc{\selemtwo} \Big[
    \darkcount{\detectormonit{2}}
    \\
    +\:& \Big(
     \presifalitoconc{0}{\selemone}
    + \presifalitoconc{1}{\selemtwo}
    \\
    -\:& 2 \presifalitoconc{0}{\selemone} \presifalitoconc{1}{\selemtwo}
    \Big)\!\left( 1 - \darkcount{\detectormonit{2}} \right)\!
    \Big].
\end{split}\end{equation}

Putting all together, we have that the starting points of the recursions and the recursive expressions of the required quantities are given by
\begin{widetext}
\begin{equation}\begin{split}
    \faveofvisclks{1}{\selemtwo\selemone}(2|0) =\:&
    \paliifres{\selemone}{0} \Big[
    \presifconc{0} \paliifres{\selemtwo}{0}
    + \presifconc{1} \paliifres{\selemtwo}{1} \darkcount{\detectormonit{1}}
    + \presifconc{2} \paliifres{\selemtwo}{2} \darkcount{\detectormonit{1}}
    \Big],
    \\
    \faveofvisclks{1}{\selemtwo\selemone}(2|1) =\:&
    \paliifres{\selemone}{1} \Big[
    \presifconc{0} \paliifres{\selemtwo}{0}
    + \presifconc{1} \paliifres{\selemtwo}{1}
    + \presifconc{2} \paliifres{\selemtwo}{2} \darkcount{\detectormonit{1}}
    \Big],
    \\
    \faveofvisclks{2}{\selemtwo\selemone}(2|0) =\:&
    \paliifres{\selemone}{0} \Big[
    \presifconc{0} \paliifres{\selemtwo}{0}
    + \presifconc{1} \paliifres{\selemtwo}{1} \darkcount{\detectormonit{2}}
    + \presifconc{2} \paliifres{\selemtwo}{2} \darkcount{\detectormonit{2}}
    \Big],
    \\
    \faveofvisclks{2}{\selemtwo\selemone}(2|1) =\:&
    \paliifres{\selemone}{1} \Big[
    \presifconc{0} \paliifres{\selemtwo}{0} \darkcount{\detectormonit{2}}
    + \presifconc{1} \paliifres{\selemtwo}{1}
    + \presifconc{2} \paliifres{\selemtwo}{2} \darkcount{\detectormonit{2}}
    \Big],
    \\
    \faveofvisclks{1}{\selemtwo\selemone}(0) =\:&
    \faveofvisclks{2}{\selemtwo\selemone}(0) =
    0.
\end{split}\end{equation}
and
\begin{equation}\begin{split}
    \faveofvisclks{1}{\selemtwo\selemone}(k|0) =\:&
    \presifconc{0} \Big\{
    \paliifres{\selemone}{0} \paliifconc{\selemtwo} \Big[
    1 - \presifalitoconc{1}{\selemtwo} \left( 1 - \darkcount{\detectormonit{1}} \right) \Big]
    + \paliifconc{\selemone} \paliifres{\selemtwo}{0}
    + \pcohclick{1} (k-3)
    \Big\}
    \\
    +&
    \presifconc{1} \Big\{
    \paliifres{\selemone}{0} \paliifconc{\selemtwo} \Big[
    1 - \presifalitoconc{1}{\selemtwo} \left( 1 - \darkcount{\detectormonit{1}} \right) \Big]
    \\
    +& \paliifconc{\selemone} \paliifres{\selemtwo}{1} \Big[
    1 - \presifalitoconc{0}{\selemone} \left( 1 - \darkcount{\detectormonit{1}} \right)
    \Big]
    + \pcohclick{1} (k-3)
    \Big\}
    \\
    +&
    \presifconc{2} \left[
    \faveofvisclks{1}{\selemtwo\selemone}(k-1|0)
    + \paliifconc{\selemone} \paliifres{\selemtwo}{2} \darkcount{\detectormonit{1}}
    \right],
    \\ 
    \faveofvisclks{1}{\selemtwo\selemone}(k|1) =\:&
    \presifconc{0} \left[
    \paliifres{\selemone}{1} \paliifconc{\selemtwo}
    + \paliifconc{\selemone} \paliifres{\selemtwo}{0}
    + \pcohclick{1} (k-3)
    \right]
    \\
    +&
    \presifconc{1} \Big\{
    \paliifres{\selemone}{1} \paliifconc{\selemtwo}
    + \paliifconc{\selemone} \paliifres{\selemtwo}{1} \left[ 1 - \presifalitoconc{0}{\selemone} \left( 1 - \darkcount{\detectormonit{1}} \right) \right]
    + \pcohclick{1} (k-3)
    \Big\}
    \\
    +&
    \presifconc{2} \left[
    \faveofvisclks{1}{\selemtwo\selemone}(k-1|1)
    + \paliifconc{\selemone} \paliifres{\selemtwo}{2} \darkcount{\detectormonit{1}}
    \right],
    \\ 
    \faveofvisclks{1}{\selemtwo\selemone}(k|2) =\:&
    \faveofvisclks{1}{\selemtwo\selemone}(k-1) + \paliifres{\selemone}{2} \paliifconc{\selemtwo} \darkcount{\detectormonit{1}},
\end{split}\end{equation}
and
\begin{equation}\begin{split}
    \faveofvisclks{2}{\selemtwo\selemone}(k|0) =\:&
    \presifconc{0} \Big\{
    \paliifres{\selemone}{0} \paliifconc{\selemtwo} \left[ 1 - \presifalitoconc{1}{\selemtwo} \left( 1 - \darkcount{\detectormonit{2}} \right) \right]
    \\
    +&
    \paliifconc{\selemone} \paliifres{\selemtwo}{0} \left[ \darkcount{\detectormonit{2}} + \presifalitoconc{0}{\selemone} \left( 1 - \darkcount{\detectormonit{2}} \right) \right]
    + \pcohclick{2} (k-3)
    \Big\}
    \\
    +&
    \presifconc{1} \Big\{
    \paliifres{\selemone}{0} \paliifconc{\selemtwo} \left[ 1 - \presifalitoconc{1}{\selemtwo} \left( 1 - \darkcount{\detectormonit{2}} \right) \right]
    \\
    +&
    \paliifconc{\selemone} \paliifres{\selemtwo}{1} \left[ 1 - \presifalitoconc{0}{\selemone} \left( 1 - \darkcount{\detectormonit{2}} \right) \right]
    + \pcohclick{2} (k-3)
    \Big\}
    \\
    +&
    \presifconc{2} \left[
    \faveofvisclks{2}{\selemtwo\selemone}(k-1|0)
    + \paliifconc{\selemone} \paliifres{\selemtwo}{2} \darkcount{\detectormonit{2}}
    \right],
    \\ 
    \faveofvisclks{2}{\selemtwo\selemone}(k|1) =\:&
    \presifconc{0} \Big\{
    \paliifres{\selemone}{1} \paliifconc{\selemtwo} \left[ \darkcount{\detectormonit{2}} + \presifalitoconc{1}{\selemtwo} \left( 1 - \darkcount{\detectormonit{2}} \right) \right]
    \\
    +&
    \paliifconc{\selemone} \paliifres{\selemtwo}{0} \left[ \darkcount{\detectormonit{2}} + \presifalitoconc{0}{\selemone} \left( 1 - \darkcount{\detectormonit{2}} \right) \right]
    + \pcohclick{2} (k-3)
    \Big\}
    \\
    +&
    \presifconc{1} \Big\{
    \paliifres{\selemone}{1} \paliifconc{\selemtwo} \left[ \darkcount{\detectormonit{2}} + \presifalitoconc{1}{\selemtwo} \left( 1 - \darkcount{\detectormonit{2}} \right) \right]
    \\
    +&
    \paliifconc{\selemone} \paliifres{\selemtwo}{1} \left[ 1 - \presifalitoconc{0}{\selemone} \left( 1 - \darkcount{\detectormonit{2}} \right) \right]
    + \pcohclick{2} (k-3)
    \Big\}
    \\
    +&
    \presifconc{2} \left[
    \faveofvisclks{2}{\selemtwo\selemone}(k-1|1)
    + \paliifconc{\selemone} \paliifres{\selemtwo}{2} \darkcount{\detectormonit{2}}
    \right],
    \\ 
    \faveofvisclks{2}{\selemtwo\selemone}(k|2) =\:&
    \faveofvisclks{2}{\selemtwo\selemone}(k-1) + \paliifres{\selemone}{2} \paliifconc{\selemtwo} \darkcount{\detectormonit{2}},
\end{split}\end{equation}
\end{widetext}

\section{Alternative USD measurement}\label{sec:USDM2}

In this Appendix we introduce an alternative USD setup for Eve that, unlike the USD1, allows her to identify the decoy signals $\cowsig{2}$, but provides a lower overall success probability than USD1.
We call this scheme USD2 and, as we did for the USD1, below we account for the most common imperfections in its model, and calculate its corresponding measurement statistics.

\begin{figure}
  \includegraphics[width=1\columnwidth]{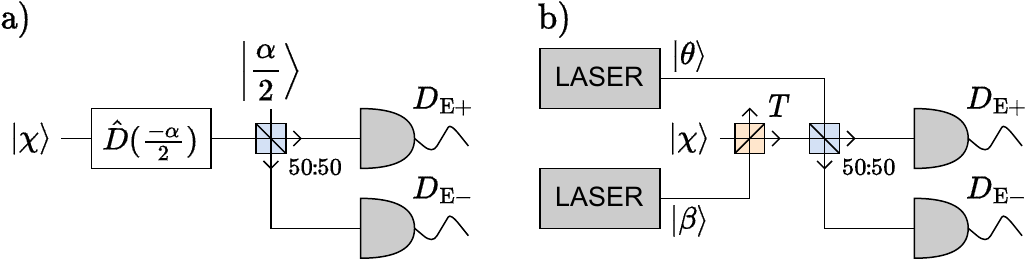}
  \caption{Alternative USD measurement setup for Eve, which we refer to as USD2.
  a) Ideal optical model.
  b) Proposed experimental realization. The values of $\theta$ and $\beta$ are given in the text.}
  \label{fig:scheme_suboptimal}
\end{figure}

The idealized optical scheme is shown in \cref{fig:scheme_suboptimal}a.
First, an optical displacement $\hat D(-\alpha/2)$ is applied to the pulses generated by Alice, transforming each pulse $\ket{\eveinput}$ into $\Ket{\eveinput-\frac{\alpha}{2}}$.
Subsequently, each displaced pulse enters a 50:50 beamsplitter, where it interferes with a coherent state $\ket{\alpha/2}$.
The resulting state $\ket{\eveinput / \sqrt{2}}$ ($\ket{(\eveinput-\alpha)/\sqrt{2}}$) at the output port of the beamsplitter associated with constructive (destructive) interference is then detected by $D_{\detectoreve[+]}$ ($D_{\detectoreve[-]}$).
Importantly, this means that, in the absence of imperfections, a click in $D_{\detectoreve[+]}$ ($D_{\detectoreve[-]}$) can only occur when Alice prepares a coherent (vacuum) pulse.

Since Alice's signals have two optical pulses, Eve has to implement this measurement twice, each time for a different pulse.
The correspondence between each measurement result $\eve_i$ and the pattern of detections needed to prompt it is shown in \Cref{tab:tabtruth_usdm2}.

\begin{table}
    \newcommand{\clickneeded}{ $\checkmark$ }
    \newcommand{\noclickneeded}{ {\sf X} }
    \newcommand{\anythingneeded}{ -- }
    \centering
    \begin{tabular}{c|c|c|c|c}
        & $D_{\detectoreve[1+]}$ & $D_{\detectoreve[1-]}$ & $D_{\detectoreve[2+]}$ & $D_{\detectoreve[2-]}$
        \\ \hline
        $\eve_0$ &
        \anythingneeded & \noclickneeded & \noclickneeded & \clickneeded
        \\ \hline
        $\eve_1$ &
        \noclickneeded & \clickneeded & \anythingneeded & \noclickneeded
        \\ \hline
        $\eve_2$ &
        \clickneeded & \noclickneeded & \clickneeded & \noclickneeded
        \\ \hline
        $\eve_3$ & \multicolumn{4}{c}{Otherwise}
    \end{tabular}
    \caption{Assignments between click patterns and measurement outcomes for USD2.
    $\deteve[1\pm]$ ($\deteve[2\pm]$) refers to the detector $\deteve[\pm]$ acting in the first (second) optical pulse of a signal.
    Symbols `\clickneeded', `\noclickneeded' and `\anythingneeded' mean that, for each measurement result, the indicated detector clicks, does not click, or is irrelevant, respectively.}
    \label{tab:tabtruth_usdm2}
\end{table}

The experimental setup for USD2 is shown in \cref{fig:scheme_suboptimal}b.
As with USD1, in practice one can approximate the optical displacement $\hat D(-\alpha/2)$ with a beamsplitter of transmittance $T\approx1$ together with an interfering offset coherent pulse $\ket{\beta}$ satisfying
\begin{equation}\label{eq:usd2_beta}
    \beta =
    -\sqrt{\frac{T}{1-T}} \frac{\alpha}{2}.
\end{equation}

With this choice of $\beta$, Eve's approximated displacement operation performs the transformation $\ket{0}\to\ket{-\sqrt{T}\alpha/2}$ and $\ket{\alpha}\to\ket{\sqrt{T}\alpha/2}$.
Since this introduces some loss, we adjust the amplitude of the interfering signal $\ket{\theta}$ at the second beamsplitter to
\begin{equation}\label{eq:usd2_theta}
    \theta =
    \sqrt{T} \frac{\alpha}{2}.
\end{equation}

To fairly compare the performance of USD2 with that of USD1 in realistic scenarios, we consider the same imperfections in both setups.
This includes the possibility of a small phase shift $\phi$ of the incoming pulses $\ket{\eveinput}$, the use of imperfect detectors with detection efficiency $\effeve$ and dark-count probability $\darkcount{\detevelabel}$, intensity fluctuations in the coherent pulses $\ket{\beta}$ and $\ket{\theta}$, and, lastly, a potentially imperfect mode overlap at each of the beamsplitters.
To characterize this latter effect we use again the simple model from~\cite{Laiho_Probing:2009}.
Moreover, for simplicity, we assume that the percentage of deviation in intensity is equal for $\ket{\beta}$ and $\ket{\theta}$, and modifies their values, respectively, by
\begin{equation}\label{eq:usd2_interfering}
    \sigma = \sqrt{1 + \amperror} \beta,
    \qquad
    \varsigma = \sqrt{1 + \amperror} \theta,
\end{equation}
where $\amperror \!\in\! [-1,\infty)$.

\begin{figure}
    \centering
    \includegraphics[width=1\columnwidth]{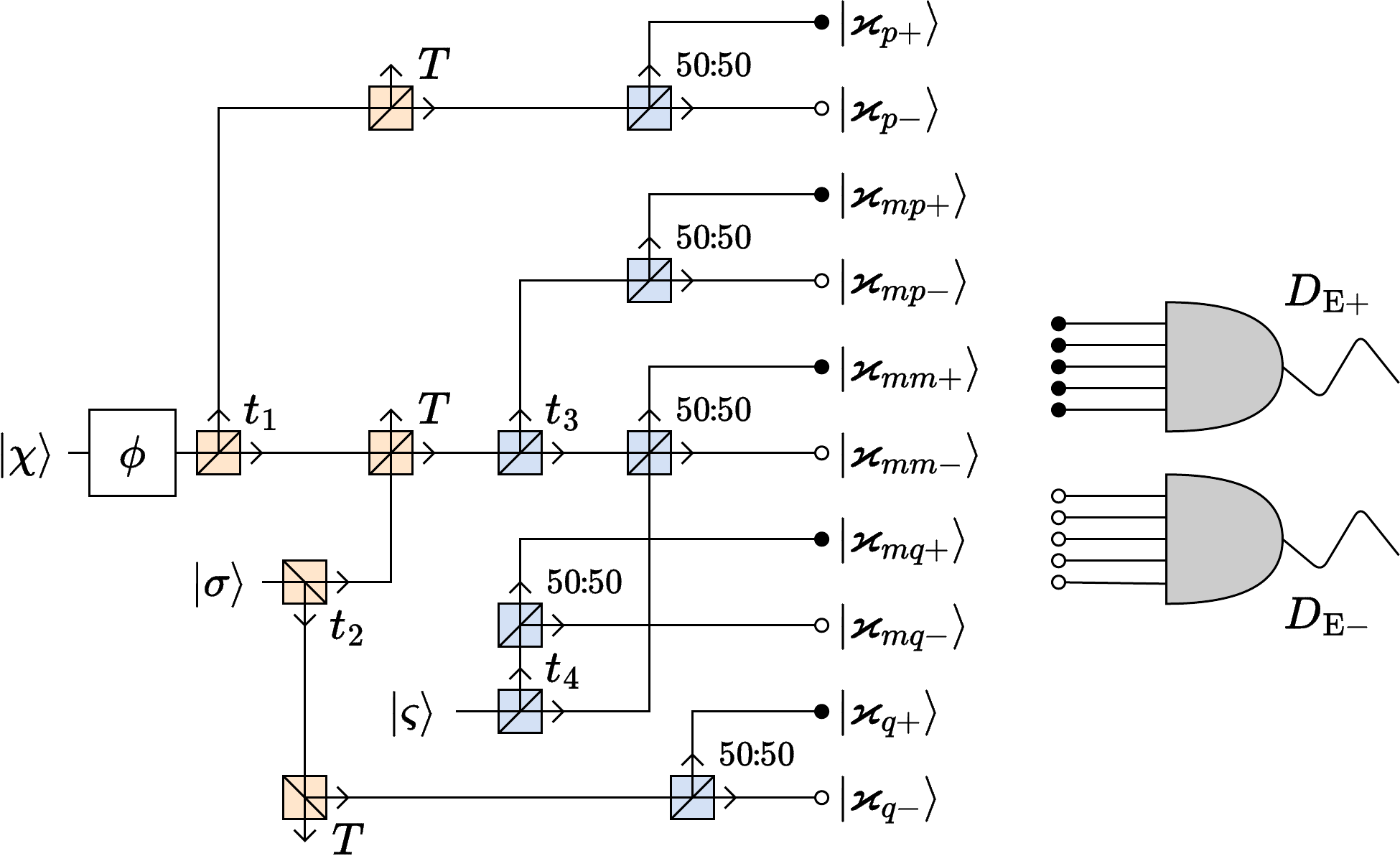}
    \caption{Model of the alternative USD measurement, USD2, including the main experimental imperfections.
    The beamsplitters depicted in orange (blue) are those related to the behaviour of the first (second) beamsplitter of the experimental setup.
    Each output mode of the scheme marked with a black (white) circle is received by the detector $\deteve[+]$ ($\deteve[-]$). That is, the total intensity received by each detector is the sum of the intensities in each of its corresponding input modes.}
    \label{fig:model_suboptimal}
\end{figure}

The complete model with imperfections is depicted in \cref{fig:model_suboptimal}.
Similar to the analysis of USD1, the parameter $t_1$ ($t_2$) denotes here the fraction of $\ket{\eveinput}$ ($\ket{\sigma}$) that correctly interferes at the first beamsplitter.
Similarly, for the 50:50 beamsplitter, $t_3$ ($t_4$) represents the fraction of light in the first (second) input port of this beamsplitter that correctly interferes.
For simplicity, we disregard any interference effect in the second beamsplitter between those optical modes that did not correctly interfere in the first one.

The total intensity received by $\deteve[+]$ ($\deteve[-]$) given that Alice sends $\ket{\eveinput}$, namely $\intenseve[+]$ ($\intenseve[-]$), is the sum of the intensities at each independent optical mode within the constructive (destructive) output port of the second beamsplitter. That is, $\intenseve[\pm] = \sum_{x} |\varkappa_{x\pm}|^2$, where $\ket{\varkappa_{x\pm}}$ denotes the equivalent coherent state in the output mode $x\pm$, with $x \in \{p,q,mp,mm,mq\}$ (see \cref{fig:model_suboptimal}).
Therefore, we have that
\begin{equation}\label{eq:intenseve_usd2}\begin{split}
    \intenseve[\pm] =\:&
    \frac{T}{4} \Big[
    2|\eveinput|^2
    + \left( 1 + \amperror \right) |\alpha|^2 \left( 1 \mp \sqrt{t_2t_3t_4} \right)
    \\
    -\:& 2\sqrt{t_1\left( 1 + \amperror \right)}\left( \sqrt{t_2} \!\mp\! \sqrt{t_3t_4} \right) \textrm{Re}\{\chi \alpha^* e^{i\phi}\}
    \Big].
\end{split}\end{equation}

Let $\pnceveifvac[\pm]$ ($\pnceveifcoh[\pm]$) be the probability that $\deteve[\pm]$ does not click given that Alice prepares $\ket{0}$ ($\ket{\alpha}$).
This probabilities can be straightforwardly computed from $\pnceve[\pm]= (1-\darkcount{\detevelabel}) \exp\{-\effeve\intenseve[\pm]\}$.
Then, by particularizing \cref{eq:intenseve_usd2} to each possible input state, we obtain
\begin{equation}\label{eq:intenseve_cond_usd2}\begin{split}
    \intenseveifvac[\pm] =\:&
    \frac{T}{4} |\alpha|^2 \left( 1 + \amperror \right) \left( 1 \mp \sqrt{t_2t_3t_4} \right),
    \\
    \intenseveifcoh[\pm] =\:&
    \frac{T}{4} |\alpha|^2 \Big[
    2 + \left( 1 + \amperror \right) \left( 1 \mp \sqrt{t_2t_3t_4} \right)
    \\
    -\:& 2\sqrt{t_1\left( 1 + \amperror \right)} \left( \sqrt{t_2} \mp \sqrt{t_3t_4} \right) \cos\phi
    \Big].
\end{split}\end{equation}

Finally, given the assignments presented in \cref{tab:tabtruth_usdm2}, it is immediate to calculate the probabilities $\presifali{i}{j}$ for the setup USD2 as
\begin{equation}\begin{split}
    \presifali{0}{0} =\:&
    \presifali{1}{1} =
    \\
    &\pnceveifcoh[-]
    \pnceveifvac[+]
    \left( 1 - \pnceveifvac[-] \right),
    \\
    \presifali{0}{1} =\:&
    \presifali{1}{0} =
    \\
    &\pnceveifvac[-]
    \pnceveifcoh[+]
    \left( 1 - \pnceveifcoh[-] \right),
    \\
    \presifali{0}{2} =\:&
    \presifali{1}{2} =
    \\
    &\pnceveifcoh[-]
    \pnceveifcoh[+]
    \left( 1 - \pnceveifcoh[-] \right),
    \\
    \presifali{2}{0} =\:&
    \presifali{2}{1} =
    \\
    &\pnceveifcoh[-]
    \pnceveifvac[-]
    \\
    \times\:&
    \left( 1 - \pnceveifcoh[+] \right)
    \left( 1 - \pnceveifvac[+] \right),
    \\
    \presifali{2}{2} =\:&
    \pnceveifcoh[-]^2
    \left( 1 - \pnceveifcoh[+] \right)^2,
    \\
    \presifali{3}{i} =\:&
    1 - \sum_{j=0}^2 \presifali{j}{i}
    \quad
    \textrm{for } i \in \{0,1,2\}.
\end{split}\end{equation}

\section{Effect of other imperfections besides mode mismatch}\label{sec:other_imperfections}

As mentioned in the main text, the crucial imperfection that determines the success of Eve's attack is the mode mismatch at her beamsplitters.
To show this, we examine here how the remaining protocol and experimental parameters affect the expected value of the metrics in the presence of the attack.
In particular, we fix all these parameters to the values used in the main text (see \cref{tab:params}) with the exception of the parameter we want to study in each particular case.
Besides, as done in the main text, the parameters that quantify the quality of the mode overlap at Eve's beamsplitters are set to $t_1=t_2=t_3=t_4=1-\varepsilon$, where here we pick $\varepsilon=2\cdot10^{-3}$.
This value of $\varepsilon$ sufficiently low to ensure that both USD1 and USD2 succeed if all the other parameters are fixed to the values shown in \cref{tab:params}.
Once again, for simplicity, we consider the same dark-count probability $\darkcount{}$ for all detectors, \textit{i.e.}, $\darkcount{\detectoreve}= \darkcount{\detectordata}= \darkcount{\detectormonit{X}}=\darkcount{}$, with $X\!\in\!\set{1,2}$.

\subsection{Effect of the intensity of Alice's pulses}
\begin{figure}
    \centering
    \includegraphics[width=1\columnwidth]{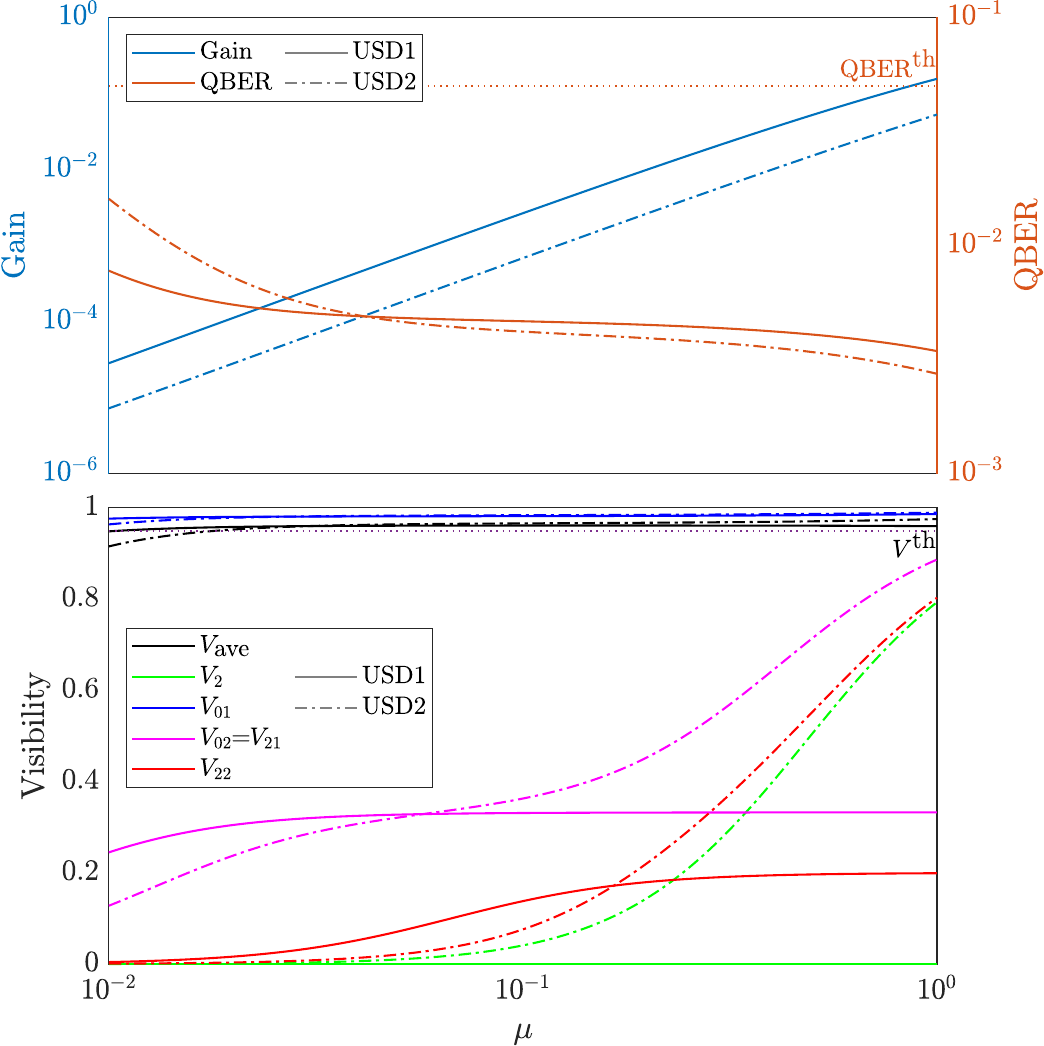}
    \caption{Gain, QBER and visibilities vs the intensity of Alice's pulses $\mu=|\alpha|^2$.
    }
    \label{fig:overMu}
\end{figure}

First, we investigate the influence of the intensity $\mu=|\alpha|^2$ of Alice's pulses on Eve's attack performance. This is shown in \cref{fig:overMu}, where the expected values of the metrics as a function of the intensity $\mu$ are displayed. Not surprisingly, the gain rises in alignment with $\mu$, as the probability of a conclusive measurement rises as well. The QBER and visibilities also improve as $\mu$ increases, as a higher gain reduces the impact of dark counts on Bob's detectors on these metrics. Notably, the QBER stays well below the critical threshold for all evaluated intensities.

Concerning the visibilities, we note that those computed for USD2 show a continuous growth with $\mu$ accross the entire considered range. This is because this scheme can identify the three signals emitted by Alice, and the probability of a conclusive measurement increases with $\mu$. However, for USD1, the visibility of sequences containing decoy signals only improves at low $\mu$, stabilizing once the impact of the dark counts at Bob's detector becomes negligible. In particular, we note that the visibilities $\visibility{02}$ and $\visibility{21}$ reach this regime quicker than $\visibility{22}$, as the sequence ``22" is more frequently resent as a vacuum signal.

\subsection{Effect of the maximum length of the conclusive block, $\maxblocklen$}

Next we focus on the maximum length of a conclusive block, $\maxblocklen$, which serves Eve to cap the memory resources required to record all the measurement results, as well as the time delay she has to introduce in the channel to apply her block processing strategy.

\begin{figure}
    \centering
    \includegraphics[width=1\columnwidth]{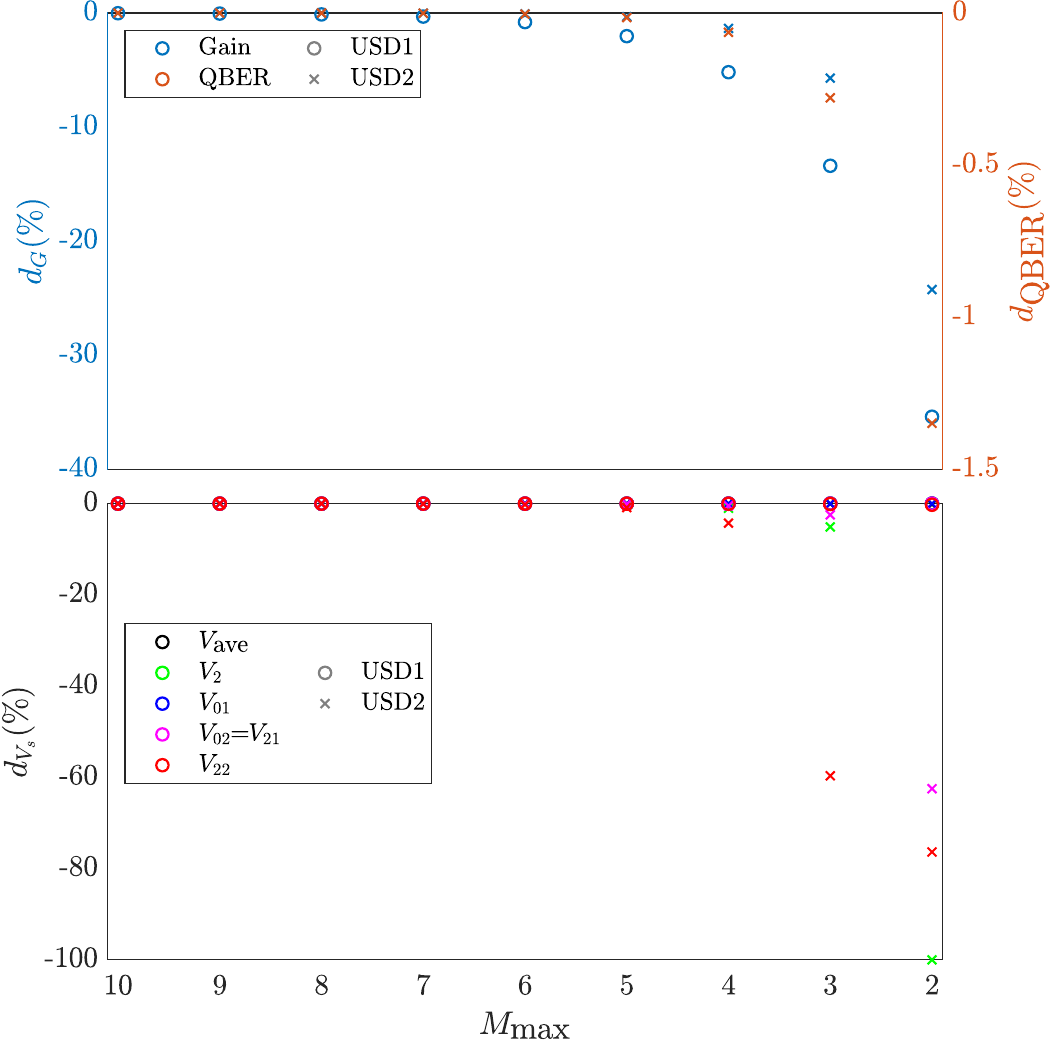}
    \caption{Relative difference of the gain ($d_{\gain}$), QBER ($d_{\qber}$) and visibilities ($d_{\visibility{s}}$), in percentage, for different values of $\maxblocklen$ when compared to the same metrics with $\maxblocklen\to\infty$, computed as in \cref{eq:relative_diff_mmax}. The intensity has been set to $\mu=1$ and the case $\maxblocklen\to\infty$ is approximated by $\maxblocklen=10^6$.\label{fig:overMmax}}
\end{figure}

Previous analysis in \cite{Gonzalez_Bounds:2020,Trenyi_Attack:2021} asserted that the variation of the metrics with $\maxblocklen$ is negligible for values of $\maxblocklen>10$ when considering practical values of $\alpha$.
To confirm this, we plot in \cref{fig:overMmax} the relative difference $d_m(\maxblocklen)$ with respect to each metric $m \in \set{\gain, \qber, \visibility{s}}$ for several values of $\maxblocklen$, where
\begin{equation}\label{eq:relative_diff_mmax}
    d_m(\maxblocklen):= \frac{m(\maxblocklen)-m(\infty)}{m(\infty)},
\end{equation}
and $m(\maxblocklen)$ is defined as the value of the metric $m$ if Eve's blocks are limited to $\maxblocklen$ pulses, while $m(\infty)$ represents the value of that metric when there is no limit to the length of Eve's blocks (we approximate this by setting $\maxblocklen=10^6$).

The results shown in \cref{fig:overMmax} are obtained considering $\mu=1$ (that is, $|\alpha|=1$), since this is close to the upper end of the practical values used in realistic COW implementations.
It can be seen that the gain loses around a 35\% of its value for USD1 when setting $\maxblocklen=2$, and around 25\% for USD2.
Nevertheless, the relative difference for values of $\maxblocklen>8$ is near zero, showing that $\maxblocklen=10$ indeed attains a similar performance as the asymptotic case, in terms of the gain.

On the other hand, the value of the QBER only varies up to $\approx1\%$ of its value when $\maxblocklen=2$, rapidly approaching zero difference when $\maxblocklen$ increases, which proves that this metric does not degrade a lot when using short blocks.
This is to be expected, as the actual probability of Eve introducing errors does not change for longer blocks, so the variation comes only from a lower significance of the effect of dark counts in Bob.

Lastly, we see that the change in the visibilities over $\maxblocklen$ is quite small for USD1, and indeed the scale of this variation is similar to the one observed for the QBER.
This is because this variation stems, once again, from an increase of dark counts in Bob when Eve sends only short blocks.
However, the visibilities in USD2 do degrade significantly when the length of the blocks is more limited.
In particular, the visibilities that deal with the decoy signal are the most affected, with $\visibility{2}$ degrading up to 100\% when $\maxblocklen=2$.
The reason is that shorter blocks impose an artificial bias against the retransmission of decoy signals by Eve, even when properly identified, due to her processing strategy.
In the limit of $\maxblocklen=2$, in fact, Eve never resends $\cowsig{2}$, as this signal is translated into $\vacsig$ when it is placed at the edge of a block.
Thus, the decreased probability of resending a decoy signal makes the behaviour of the USD2 similar to that observed in previous sections, where the blocks were short due to the small intensity, and therefore most of the signals sent to Bob when Alice sends $\cowsig{2}$ correspond to Eve's misidentification of $\cowsig{2}$ by one of the data signals, which naturally leads to poor visibility results.

In any case, it can be highlighted from \cref{fig:overMmax} that all metrics are very close to their asymptotic value when setting $\maxblocklen=10$, even when Alice sends pulses with a relatively high intensity, and therefore it is sufficient for Eve to use this configuration in practical scenarios.

\subsection{Effect of the transmittance of the asymmetric beamsplitter, $T$}
Now we investigate the performance of Eve's attack as a function of the transmittance $T$ of her asymmetric beamsplitter, which she uses to approximates an optical displacement in both USD measurement schemes. While this approximation is only accurate when $T\approx 1$, the simulations indicate that the value of $T$ has a minimal impact on the attack's feasibility.

\begin{figure}
    \centering
    \includegraphics[width=1\columnwidth]{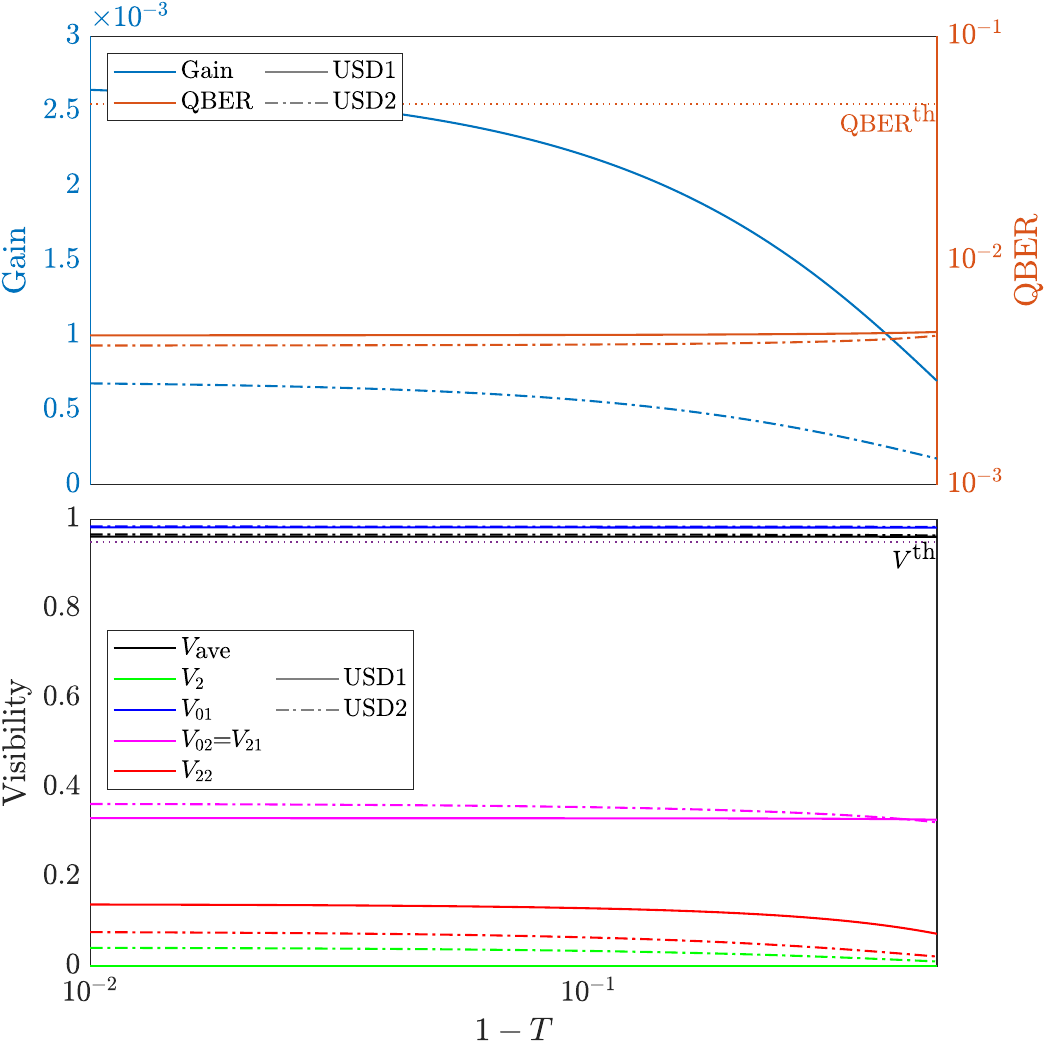}
    \caption{Gain, QBER and visibilities vs the transmittance of Eve's asymmetric beamsplitter, $T$.}
    \label{fig:overT}
\end{figure}

This is illustrated in Figs. \ref{fig:overT}a and \ref{fig:overT}b, where the metrics are plotted against $1-T$, within the range $(1-T)\in[10^{-3},0.5]$. The simulations indicate that, as $T$ decreases, so does the expected gain in both schemes, which is attributed to the small loss introduced by the approximate displacement. Importantly, however, the QBER and visibilities are largely unaffected by changes in $T$ within a reasonable range, and both the QBER and $\visave$ remain acceptable even at $T=0.5$. This stability is due to our specific choice of $\beta$.

\begin{figure}
    \centering
    \includegraphics[width=1\columnwidth]{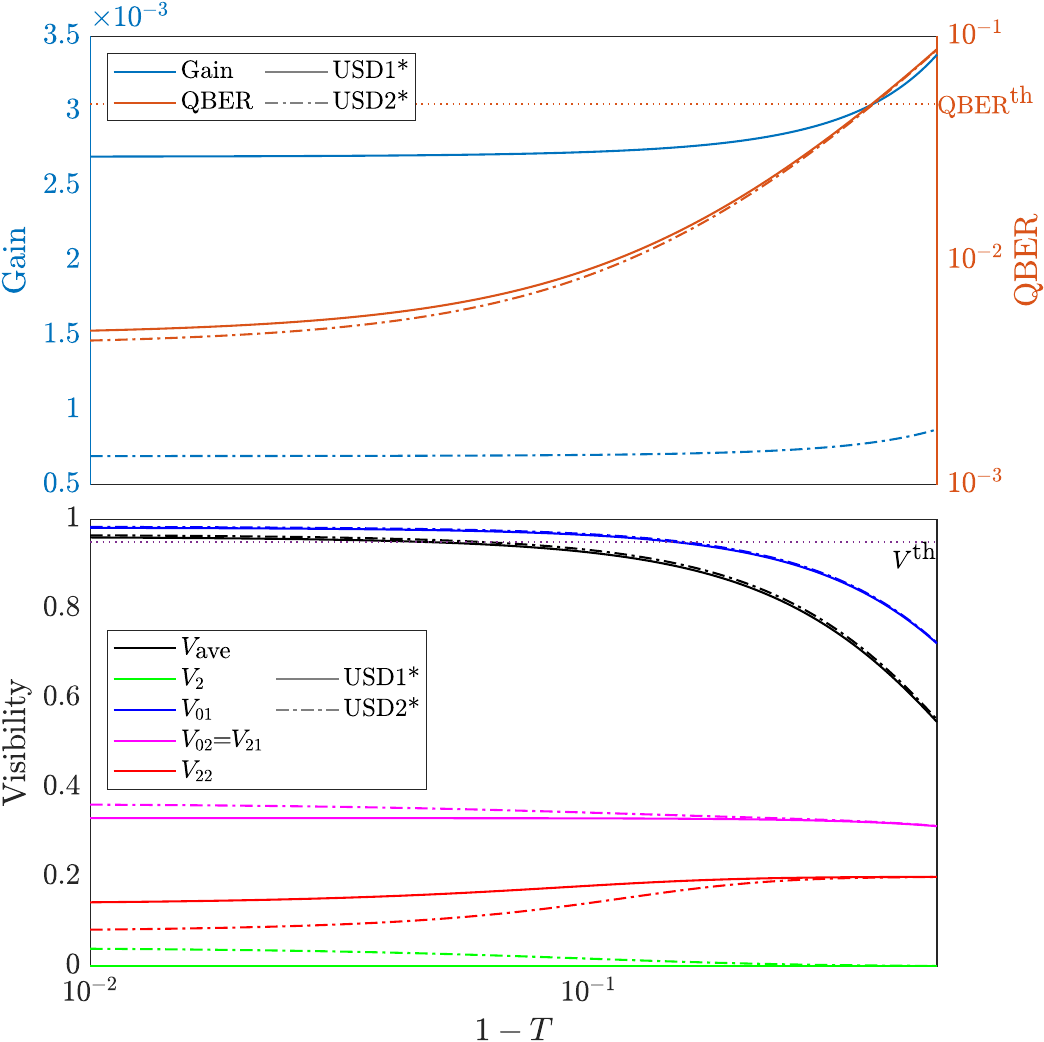}
    \caption{Gain, QBER and visibilities vs the transmittance of Eve's asymmetric beamsplitter, $T$, for the alternative schemes USD1* and USD2*.
    In particular, we consider that $\beta=-\frac{\alpha}{\sqrt{1-T}}$ ($\beta=-\frac{\alpha}{2\sqrt{1-T}}$) for USD1* (USD2*).}
    \label{fig:overT_alt}
\end{figure}

Indeed, one could alternatively set $\beta=-\frac{\alpha}{\sqrt{1-T}}$ ($\beta=-\frac{\alpha}{2\sqrt{1-T}}$) for USD1 (USD2), and the resulting scheme would still approximate the desired displacement for $T\approx1$. For instance, for USD1, this leads to the transformation $\ket{0}\to\ket{-\alpha}$ and $\ket{\alpha}\to\ket{(\sqrt{T}-1)\alpha}$. This choice avoids incurring into additional losses (thus resulting in a higher gain). However, this comes at the cost of increasing the probability of erroneous clicks in Eve's detector, making it a less convenient option for her. This is illustrated in \cref{fig:overT_alt}, where we consider these alternative values of $\beta$, and call the resulting schemes as USD1* and USD2*.

\subsection{Effect of the phase shift, $\phi$}
\begin{figure}
    \centering
    \includegraphics[width=1\columnwidth]{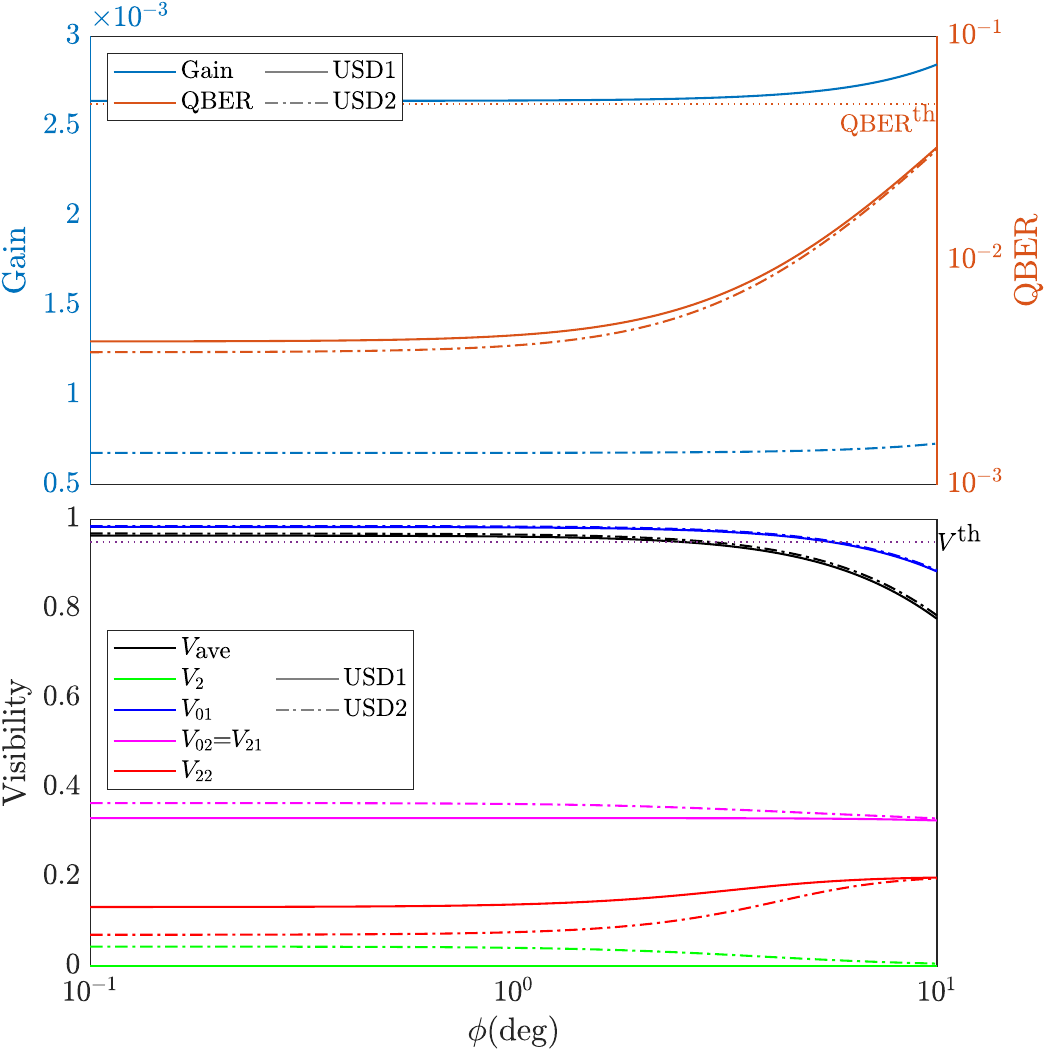}
    \caption{Gain, QBER and visibilities vs the phase shift of pulses received by Eve, $\phi$ (in degrees).}
    \label{fig:overphi}
\end{figure}

\cref{fig:overphi} shows the dependency of the expected values of the metrics on the phase shift $\phi$. Since the phase shift $\phi$ only has an impact when Alice transmits a coherent pulse, the decoy signal will be clearly influenced the most. In fact, the probability of misidentifying $\cowsig{2}$ as a data signal increases with $\phi$. As a consequence, signals that are never (in USD1) or rarely (in USD2) identified for low $\phi$, result in conclusive measurements when this parameter increases, which in turn increases the gain. Of course, the probability of misidentifying a data signal also grows, which leads to an increment of the QBER, as shown in \cref{fig:overphi}. Nevertheless, we note that the QBER remains relatively low even for phase shifts of several degrees.

Similarly, the visibilities generally decrease with $\phi$, and in this case $\visave$ does fall bellow the acceptance threshold when $\phi \gtrapprox 2.5^\circ$. Interestingly, $\visibility{22}$ grows with $\phi$ until it reaches a certain point at which it stabilizes. This is again due to the increasing probability of misidentifying $\cowsig{2}$ as one of the data signals. Note that, since the sequence ``01" is favoured by Eve's processing, the visibility of any two-signal sequence is expected to be nonzero even if the outcome of the USD measurements are totally random. In particular, in that extreme scenario, the sequence ``22" could be identified as any other possible sequence by Eve, but the only two-signal block that she resends to Bob is ``01", which always triggers the correct detector in the monitoring line, and hence increases $\visibility{22}$.

\subsection{Effect of the intensity deviation, $\amperror$}
\begin{figure}
    \centering
    \includegraphics[width=1\columnwidth]{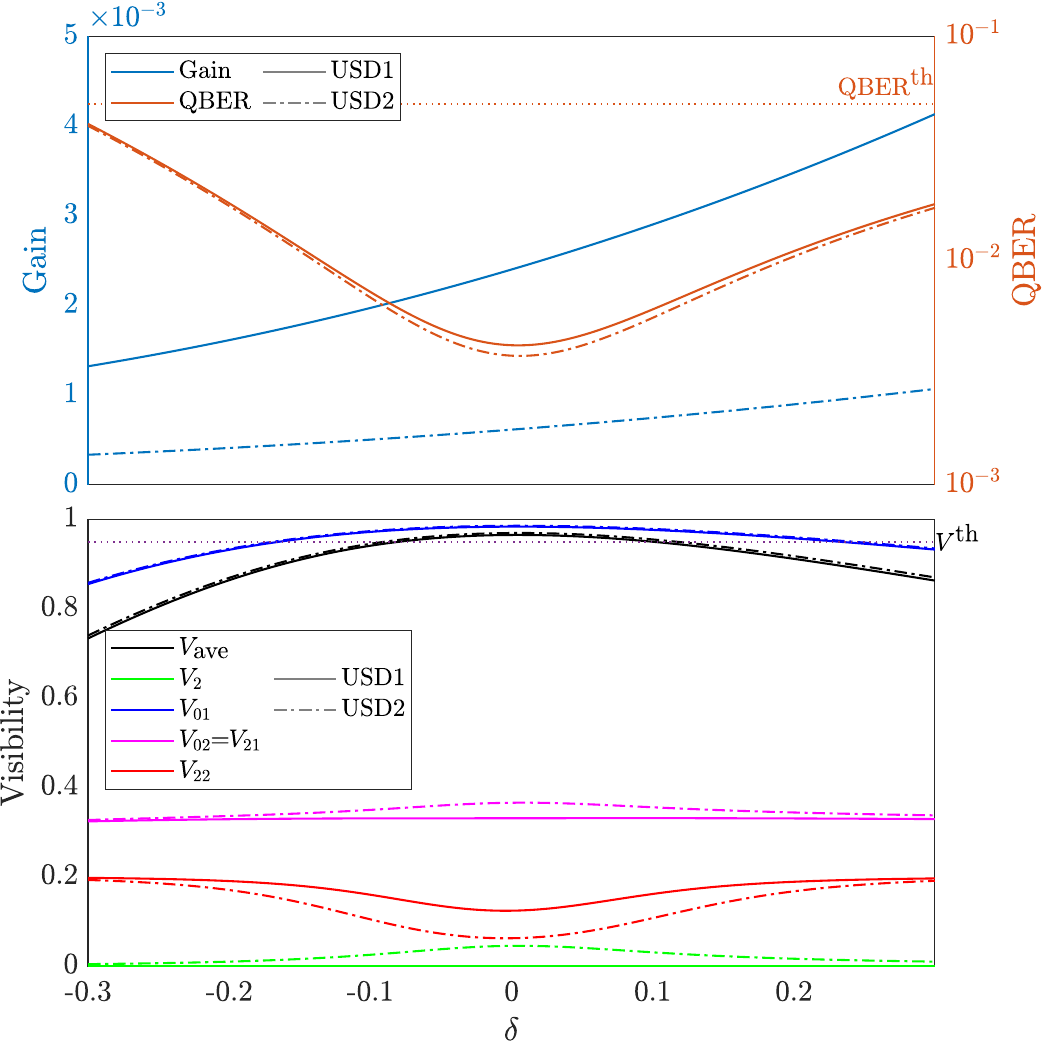}
    \caption{Gain, QBER and visibilities vs the intensity deviation of $\ket{\beta}$ (and $\ket{\theta}$ in USD2), $\amperror$.}
    \label{fig:overdelta}
\end{figure}

\cref{fig:overdelta} illustrates the impact of small deviations in the intensities of Eve's pulses, quantified with the parameter $\amperror$, on the protocol metrics.
Since $\amperror$ can take both negative and positive values, as expressed in \cref{eq:usd1_beta_err,eq:usd2_interfering}, we plot it here in the range $[-0.3,0.3]$, which corresponds to a deviation of $\pm30\%$ over the intensity of the pulses.

Interestingly, the effects of positive or negative deviations are relatively distinct.
In particular, the gain increases with $\amperror$ through the entire depicted range.
This is because higher intensities of $\ket{\sigma}$ and $\ket{\varsigma}$ result in larger click probabilities, especially in the case where Alice sends vacuum, since then Eve's signals are the only source of energy in the circuit.
Notably, the QBER remains below its corresponding threshold even for significantly high deviations, although the results are worse for negative values of $\amperror$.
This is because the probability of Eve observing a click when she is not supposed to (\textit{i.e.}, when a precise interference is intended to cancel the signal out) increases roughly as much for both positive and negative deviations.
On the other hand, the same probability given that Eve is indeed supposed to observe a click grows with larger intensities, so the effect is relatively worse for negative deviations.
Finally, \cref{fig:overdelta} shows that the visibilities are more sensitive to $\amperror$.
They exhibit a behaviour similar to that shown in~\cref{fig:metrics_visibs,fig:overphi}, with a sharper decrease in the value of the average visibility $\visave$ obtained for negative deviations, for similar reasons as for the $\qber$.
In particular, $\visave$ is above the proposed acceptance threshold for deviations in the approximate range $\amperror\in(-0.08,0.1)$, which is still a relatively large margin.

\subsection{Effect of the parameters of Eve's detectors}
\begin{figure}
    \centering
    \includegraphics[width=1\columnwidth]{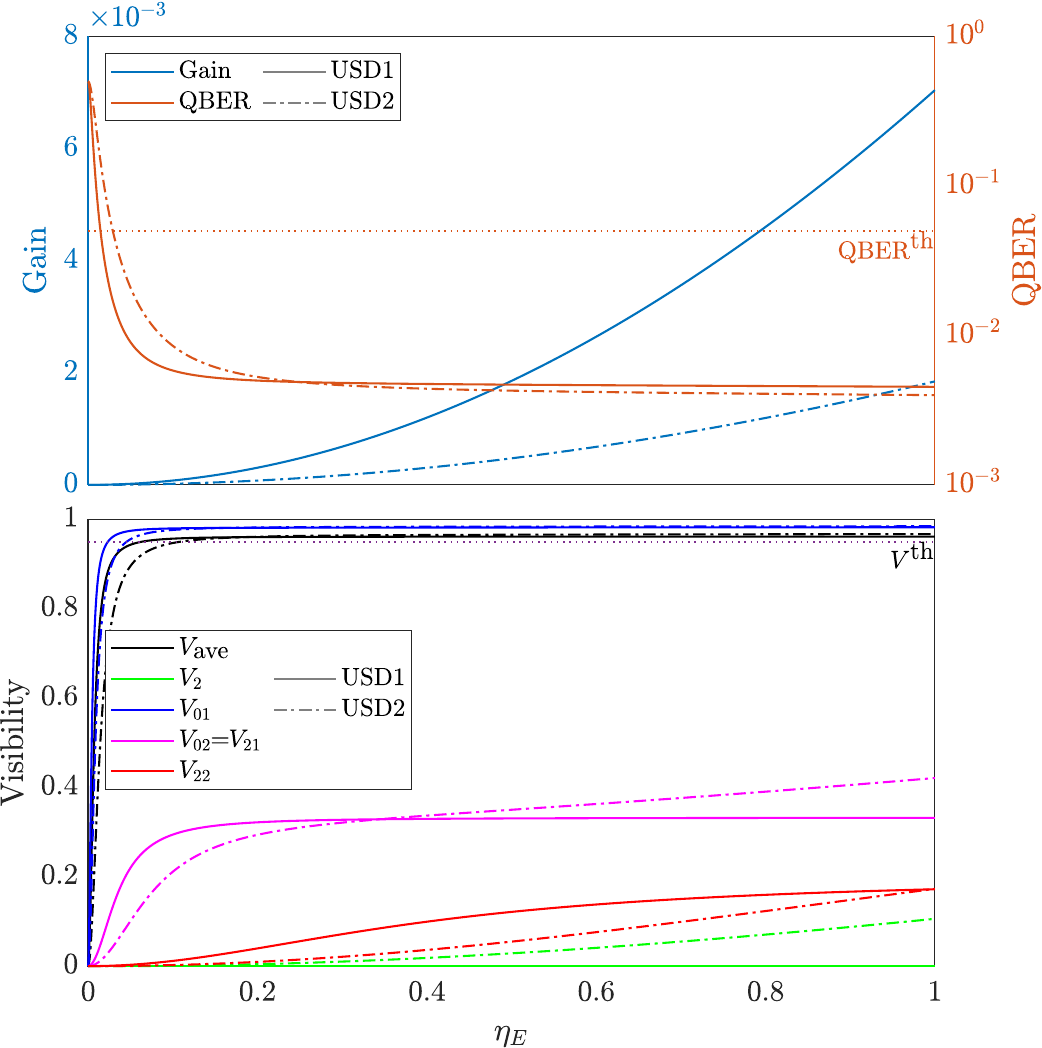}
    \caption{Gain, QBER and visibilities vs the efficiency of Eve's detectors, $\effeve$.
    }
    \label{fig:overetae}
\end{figure}

As expected, when $\effeve$ approaches zero, all metrics exhibit a degradation. This is because, in this scenario, Eve is sending a reduced number of signals to Bob, leading to a predominance of dark counts in Bob's system. This is illustrated in \cref{fig:overetae}. The remaining conclusions drawn from this figure align with those inferred from \cref{fig:overMu}. This is because the loss introduced by the nonideal detection efficiency of Eve's detector can be practically translated into an effective change in the intensity of Alice's signals to $\effeve\mu$.

\begin{figure}
    \centering
    \includegraphics[width=1\columnwidth]{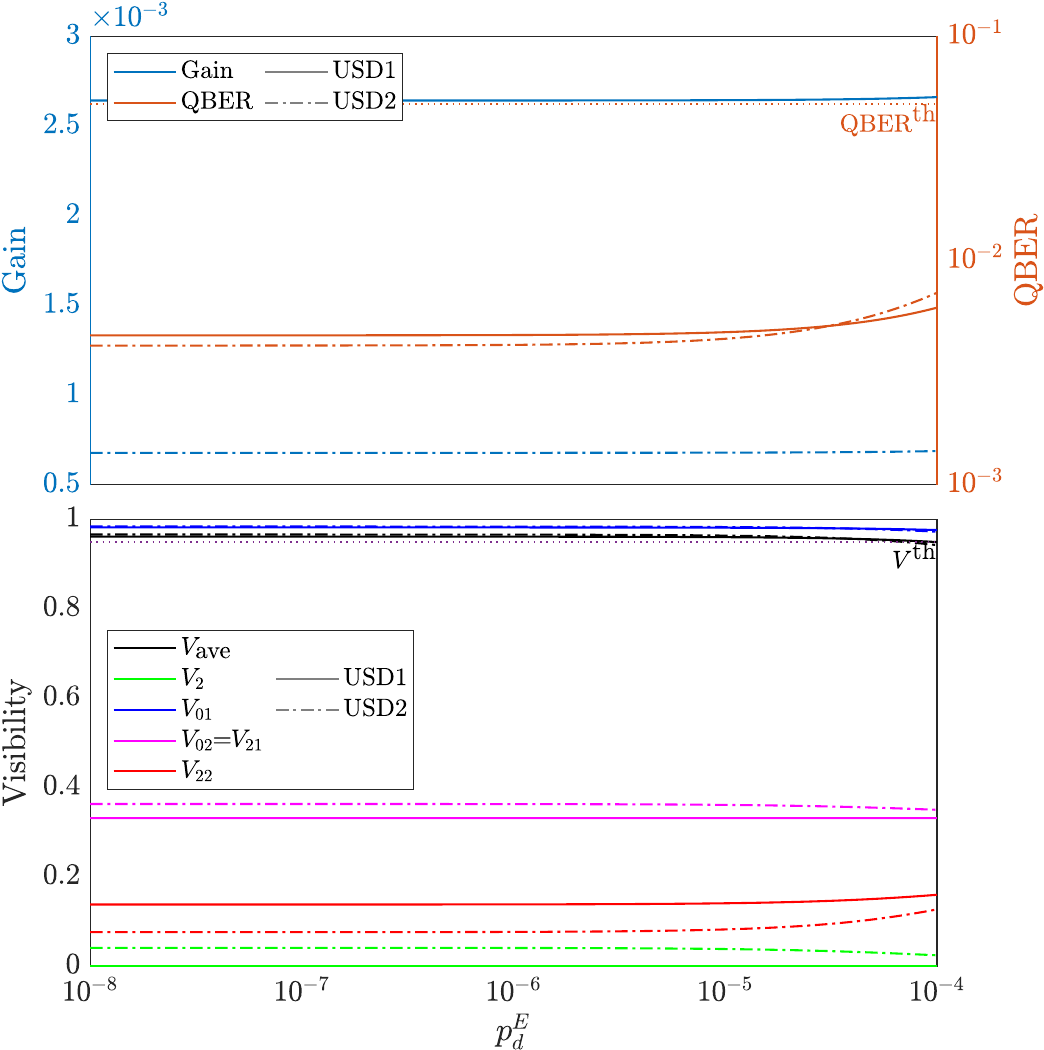}
    \caption{Gain, QBER and visibilities vs the dark-count probability of Eve's detectors, $\darkcount{\detectoreve}$. For simplicity, here we assume the same dark-count probability for all of them.}
    \label{fig:overpdEve}
\end{figure}

The dependence of the metrics on the dark-count rate $\darkcount{\detectoreve}$ of Eve's detectors is shown in \cref{fig:overpdEve}.
The gain slightly increases for high values of $\darkcount{\detectoreve}$, as more erroneous clicks in the detectors lead to more conclusive measurements. 
Naturally, this comes with a degradation of the QBER and visibilities, although this degradation is quite gentle.
In fact, the average visibility exhibits notable resilience to practical values of $\darkcount{\detectoreve}$, only falling below $\minvis\!=\!0.95$ in the approximate range $\darkcount{\detectoreve} \gtrapprox 5 \cdot 10^{-5}$.
As for the remaining visibilities, their behaviour is relatively similar to the variation with $\varepsilon$, explained more in depth in \cref{sec:explanation_visib}, albeit much more mild in magnitude.

\subsection{Effect of the parameters of Bob's detectors}
Regarding Bob's detectors, we disregard the effect of their detection efficiency, as Eve can always make them click by resending him pulses with sufficiently high intensity. Thus, we focus on the impact of the dark counts.

\begin{figure}
    \centering
    \includegraphics[width=1\columnwidth]{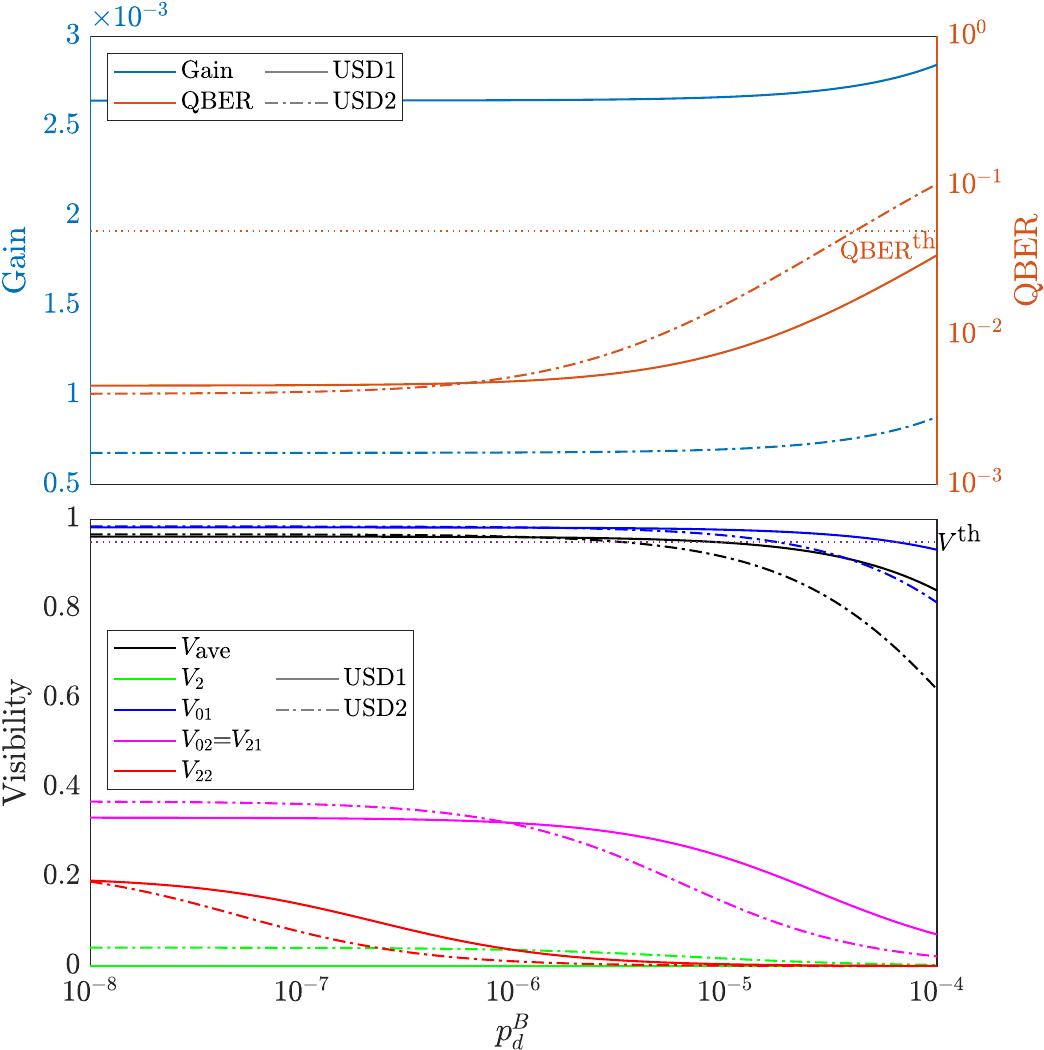}
    \caption{Gain, QBER and visibilities vs the probability of dark counts in Bob's corresponding detectors ($\darkcount{\detectordata}=\darkcount{\textrm{B}}$ for the gain and QBER, $\darkcount{\detectormonit{1}}=\darkcount{\detectormonit{2}}=\darkcount{\textrm{B}}$ for the visibilities).}
    \label{fig:overpdBob}
\end{figure}

\cref{fig:overpdBob}a shows the variation of the gain and QBER with the dark-count probability $\darkcount{\detectordata}$ at $D_{\detectordata}$.
Naturally, the gain rises with $\darkcount{\detectordata}$, as it directly increases the number of clicks at Bob's data line.
Moreover, the QBER also grows, as more random clicks lead to more errors.
Unsurprisingly, this dependence is stronger than that observed with $\darkcount{\detectoreve}$, as the dark counts at Bob's detectors more directly cause errors than those at Eve's detectors.
Indeed, for our particular choice of experimental and protocol parameters, an attack with USD2 remains viable only for $\darkcount{\detectordata} \lessapprox 4\cdot10^{-5}$, while USD1 can sustain an attack for any plotted value of $\darkcount{\detectordata}$.
Regarding the visibilities, we consider for simplicity that both detectors at the monitoring line are equal, and so $\darkcount{\detectormonit{1}}= \darkcount{\detectormonit{2}}= \darkcount{\detectormonit{X}}$.
As expected, just like the QBER, all of them get worse as $\darkcount{\detectormonit{X}}$ increases.
Still, for the parameters we consider here, the average visibility remains sufficiently high given that $\darkcount{\detectormonit{X}}\lessapprox 3\cdot10^{-6}$.

\subsection{Effect of the probability of preparing the decoy signal, $f$}
\begin{figure}
    \centering
    \includegraphics[width=1\columnwidth]{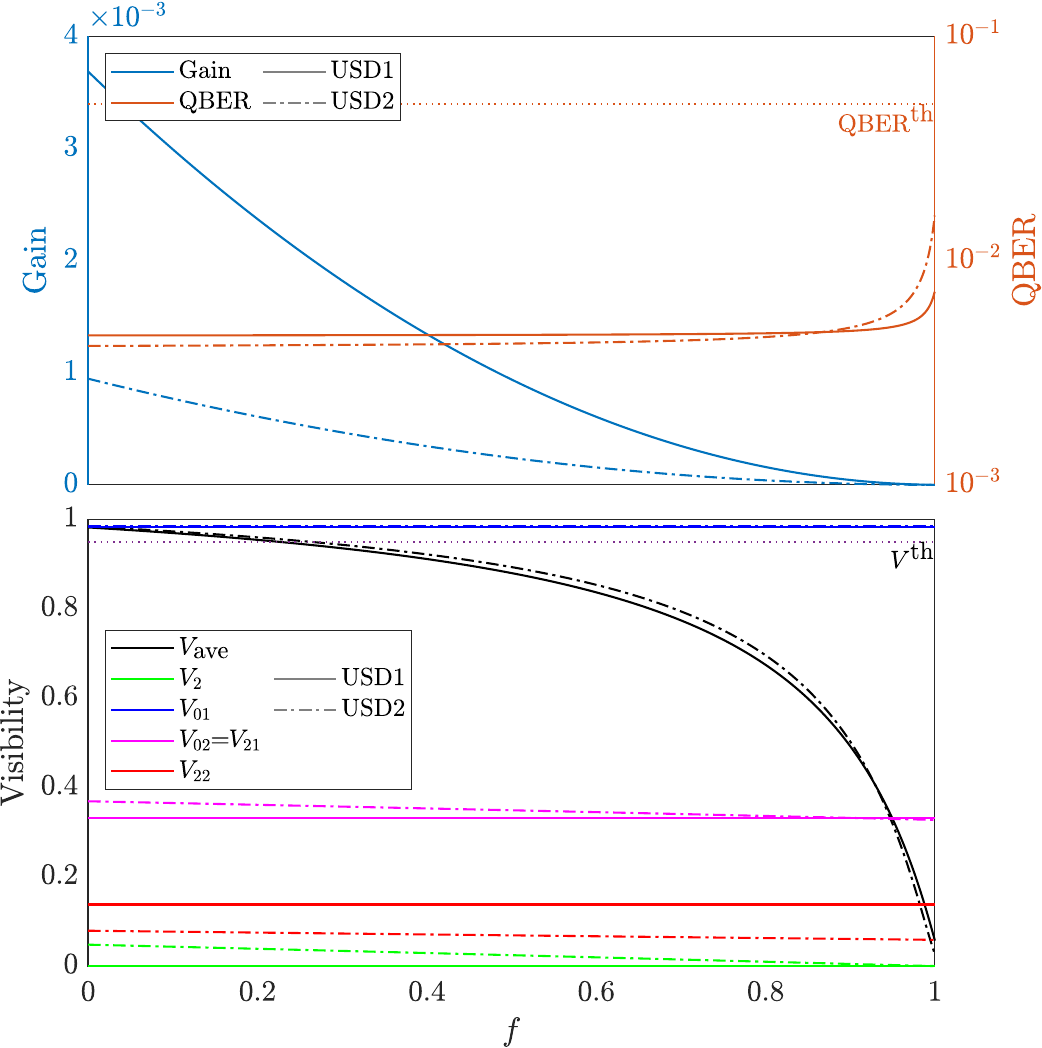}
    \caption{Gain, QBER and visibilities vs the transmission frequency of decoy signals, $f$.}
    \label{fig:overf}
\end{figure}

Finally, we investigate the effect of the decoy probability $f$ on the metrics. As shown in \cref{fig:overf}, the gain decreases to nearly zero for large values of $f$. This is not only because Eve finds it more challenging to conclusively measure the decoy signal, but also and more importantly, because Eve's processing needs data signals to be located at the edges of the blocks. Thus, a small number of data signals sent by Alice means that most conclusive measurements come from decoy signals, which cannot be resent on their own.
On the other hand, a smaller $\pconc$ also leads to a greater effect of clicks due to dark counts in Bob over the metrics, and indeed the QBER slightly grows for large values of $f$.

Regarding the visibilities, \cref{fig:overf} showcases how the average visibility falls towards 0 for increasing values of $f$.
This is because the probability of sending sequences that involve a decoy increases, which makes visibilities observing these sequences to have more weight over the final result of $\visave$.
On the other hand, the values of the individual visibilities do not decrease as significantly, and in fact, they remain essentially constant for the case of USD1.
For USD2, however, those visibilities that depend on Alice sending a decoy signal slightly decrease with $f$.
The reason for this is somewhat counter-intuitive.
As more decoy signals are sent, blocks processed by Eve become smaller, so relatively more of the decoy signals that are not turned to vacuum come from erroneously measuring them as data signals, thus leading to a decreased visibility.

In any case, selecting a high value of $f$ also severely decreases the secret-key rate of the protocol, as less data signals are emitted. Indeed, typical experiments of COW-QKD use a value of $f$ quite low, around $0.15$.

\section{Behaviour of the visibilities $\visibility{22}$, $\visibility{21}$, and $\visibility{02}$}\label{sec:explanation_visib}
When $\varepsilon$ is very small, most instances of $\cowsig{2}$ (all of them if we consider the strategy USD1) are resent to Bob as vacuum signals, so the dark counts of Bob's detectors bring the visibilities down to zero.

As $\varepsilon$ increases, so does the probability of conclusively (albeit erroneously) identifying a decoy signal $\cowsig{2}$ as $\cowsig{0}$ or $\cowsig{1}$.
In this scenario, Eve sends Bob more non-vacuum pulses, and relatively less clicks in the monitoring line are attributed to the dark counts.
In the regime where dark counts are, comparatively to signal clicks, very low, the vacuum pulses sent by Eve ---outside and at the edges of conclusive-blocks--- can be ignored.
Since USD1 never returns $\eve_2$, and USD2 has a significantly higher probability of misidentifying $\cowsig{2}$ as a data signal than correctly identifying $\cowsig{2}$ for sufficiently high $\varepsilon$, we can analyze the visibilities, in this regime, by focusing on the four possible sequences of data signals that Eve can erroneously identify: \sequence{00}, \sequence{01}, \sequence{10} and \sequence{11}.
In particular, when Bob receives \sequence{01}, only $\detbmonit[1]$ can click due to the interference of two coherent pulses.
The sequence \sequence{10} contains two vacuum pulses in the intermediate time slots, so no clicks can be observed at Bob (aside from those from dark counts).
Finally, both \sequence{00} and \sequence{11} interfere $\ket{0}$ and $\ket{\gamma}$, resulting in a click in both $\detbmonit[1]$ and $\detbmonit[2]$ with very high probability.

Let us focus now in $\visibility{22}$. We notice that, whenever Alice sends \sequence{22}, Eve misidentifies this sequence as one of the four previous sequences with equal probability, due again to the symmetry of the setup. This means that $\detbmonit[1]$ clicks with probability 3/4, while $\detbmonit[2]$ clicks with probability 2/4 (including possible double clicks in both detectors). Therefore, by using the definition of the visibility, we find that $\visibility{22} \approx 0.2$ for large values of $\varepsilon$.

Similar arguments apply to $\visibility{02}$. When Alice sends the sequence \sequence{02}, we can distinguish between two different scenarios.
If $\varepsilon$ is high enough such that the dark counts are not the main source of clicks in the monitoring line, but the probability of mistaking one data signal for another is still sufficiently low, then Eve mostly misidentifies the original sequence as \sequence{00} or \sequence{01} with equal probability, and therefore $\visibility{02} \approx 1/3$.
When $\varepsilon$ increases, misidentifications of the data signal $\cowsig{0}$ of the sequence \sequence{02} happen more often, and thus also the sequences \sequence{10} and \sequence{11}, which implies that the visibility decreases.
Analogous reasoning applies to $\visibility{21}$.

In addition, from \cref{fig:metrics_visibs} we observe that the range of values of $\varepsilon$ where dark counts are relevant is larger for USD2 than for USD1, due to the fact that USD1 offers a larger $\pconc$, and therefore more non-vacuum pulses are resent.
Moreover, this range is also considerably larger for $\visibility{22}$ than for $\visibility{02}$, since two consecutive decoy signals are less likely to yield a conclusive measurement outcome than a single one.
We also notice that USD2 performs better than USD1 for large values of $\varepsilon$.
The reason for this behavior is that the probability of erroneously identifying $\cowsig{0}$ as $\cowsig{1}$ (or viceversa), given that the measurement is conclusive, is slightly smaller in USD2.

\section{Partial attack}\label{sec:sparse_attack}

As shown in \cref{sec:results}, the average visibility $\visave$ in the presence of Eve's attack is above the acceptance threshold only for rather low values of the parameter $\varepsilon$. Nevertheless, it is still possible for Eve to remain undetected while obtaining partial information about the secret key. To this end, she can perform her attack on only a fraction of the rounds, so that the statistics from the unattacked rounds enhance the expected values of the protocol metrics, compensating for the errors introduced by her attack. Here we explain how we evaluate the expected value of the metrics when Eve executes the attack on a fraction $\attackratio$ of the rounds.

We assume that the rounds under attack are consistently clustered in large groups of consecutive rounds, allowing us to disregard any possible border effects between the unattacked and attacked signals.
Then we have that the gain is simply given by $\gain = \attackratio \gain^\attackedlabel + (1-\attackratio) \gain^\legitlabel$, where the superscript `$\attackedlabel$' indicates that it is calculated for the system that is being attacked all the protocol rounds, as described in \cref{sec:metrics}, while `$\legitlabel$' indicates that the metric is calculated in the absence of Eve.
The result of $\gain^\legitlabel$ can be computed from \cref{eq:noattack_gain}.
To calculate the QBER, on the other hand, we have to find the values of $\aveofevents{\eventdataclk}$ and $\aveofevents{\eventerror}$. Precisely, we can express these as $\aveofdataclks = \attackratio \aveofdataclks^\attackedlabel + (1-\attackratio) \aveofsigs p_\eventdataclk^\legitlabel$ and $\aveoferrors = \attackratio \aveoferrors^\attackedlabel + (1-\attackratio) \aveofsigs p_\eventerror^\legitlabel$, where
\begin{equation}\begin{split}
    p_\eventdataclk^\legitlabel =\:&
    (1-f) \left[ 1 - \left( 1 - \darkcount{\detectordata} \right)^2 e^{-\effbob\transmittance t_B |\alpha|^2} \right]
    ,
    \\ 
    p_\eventerror^\legitlabel =\:&
    (1-f) \left[ 1 + \left( 1 - \darkcount{\detectordata} \right) e^{-\effbob\transmittance t_B |\alpha|^2} \right] \frac{\darkcount{\detectordata}}{2}
    ,
\end{split}\end{equation}
are the probabilities of these events in the absence of Eve. Similarly, we can modify the values of the visibilities by making $\aveofvisclks{X}{\svector}= \attackratio \aveofvisclks{X}{\svector}^\attackedlabel + (1-\attackratio) \aveofsigs p_{\eventvisclk{X}{\svector}}^\legitlabel$, where
\begin{equation}\begin{split}
    \frac{p_{\eventvisclk{1}{2}}^\legitlabel}{f} =\:&
    \frac{p_{\eventvisclk{1}{\selemtwo\selemone}}^\legitlabel}{\palice{\selemone}\palice{\selemtwo}} =
    \\
    &1
    - \left( 1 - \darkcount{\detectormonit{1}} \right) e^{-2\effbob\transmittance (1-t_B) |\alpha|^2}
    ,
    \\ 
    \frac{p_{\eventvisclk{2}{2}}^\legitlabel}{f} =\:&
    \frac{p_{\eventvisclk{2}{\selemtwo\selemone}}^\legitlabel}{\palice{\selemone}\palice{\selemtwo}} =
    \darkcount{\detectormonit{2}},
\end{split}\end{equation}
are the probabilities corresponding to the relevant clicks when Eve does not act on the channel.

In order to compute the ratio of sifted key, $\textrm{EXT}_K$, that Eve can extract by enabling her attack during a fraction $\attackratio$ of all the communication rounds, one can observe that
\begin{equation}
    \textrm{EXT}_K =
    \frac{\attackratio \aveofevents{\eventdataclk}^\attackedlabel}
    {\attackratio \aveofevents{\eventdataclk}^\attackedlabel + (1 - \attackratio) \aveofsigs p_{\eventdataclk}^\legitlabel}.
\end{equation}
\\

\bibliography{refs}

\end{document}